\documentclass[12pt]{article}

\usepackage{epsfig}
\usepackage[utf8]{inputenc}
\usepackage{cite}
\usepackage{float}
\usepackage{hyperref}
\usepackage{epsfig}
\usepackage{amsmath}
\usepackage{amsfonts}
\usepackage{graphicx}
\usepackage{cite}
\usepackage{color}
\usepackage[dvipsnames]{xcolor}
\usepackage{multirow}
\usepackage{amssymb}

\def\be{\begin{equation}}
\def\ee{\end{equation}}

\begin{document}

\title{{\bf  \LARGE  Shadows of Exact Binary Black Holes}}

 \author{
{\large Pedro V.P. Cunha}$^{1,2}$, \
{\large Carlos A. R. Herdeiro}$^{1}$, \
{\large Maria J. Rodriguez}$^{3,4}$,  \\
\\
$^{1}${\small Departamento de F\'\i sica da Universidade de Aveiro and } \\ {\small  Centre for Research and Development  in Mathematics and Applications (CIDMA),} \\ {\small    Campus de Santiago, 3810-183 Aveiro, Portugal}
 \\
 \\
$^{2}${\small Centro de Astrof\'isica e Gravitaç\~ao - CENTRA, Departamento de F\'isica,}\\ { \small Instituto Superior T\'ecnico - IST, Universidade de Lisboa - UL,}\\ {\small Av. Rovisco Pais 1, 1049-001, Lisboa, Portugal}
 \\
 \\
$^{3}${\small Max Planck for Gravitational Physics - Albert Einstein Institute,}\\ { \small Am M\"uhlenberg 1, Potsdam 14476, Germany} 
\\
\\
  $^{4}${\small  Department of Physics, Utah State University,}\\ {\small 4415 Old Main Hill Road, UT 84322, USA}\\
}

\date{May 2018}

\maketitle

\begin{abstract}
Black hole (BH) shadows in \textit{dynamical} binary BHs (BBHs) have been produced via ray-tracing techniques on top of expensive fully non-linear numerical relativity simulations. We show that the main features of these shadows are captured by a simple {\it quasi}-static resolution of the photon orbits on top of the \textit{static} double-Schwarzschild family of solutions. Whilst the latter contains a conical singularity between the line separating the two BHs, this produces no major observable effect on the shadows, by virtue of the underlying cylindrical symmetry of the problem. This symmetry is also present in the \textit{stationary} BBH solution comprising two Kerr BHs separated by a massless strut. We produce images of the shadows of the exact stationary co-rotating (even) and counter-rotating (odd) stationary BBH configurations. This allow us to assess the impact on the binary shadows of the intrinsic spin of the BHs, contrasting it with the effect of the orbital angular momentum.
\end{abstract}

\newpage


\section{Introduction}

Bound pairs of spinning black holes (BHs) orbiting around each other, known as binary BHs (BBHs), have only very recently been observed. The first gravitational waves detected by LIGO \cite{2016PhRvL.116f1102A} confirmed the existence of BBHs in the Universe, by detecting the final stages of their inspiral and their merger. The subsequent LIGO-Virgo detections~\cite{Abbott:2016nmj,Abbott:2017vtc,Abbott:2017gyy,Abbott:2017oio} confirmed an abundant BBHs population, when one considers mildly cosmological distances. In our galaxy, on the other hand, BBH \textit{mergers} will be extremely rare events, but indirect evidence from the electromagnetic channel, supports the existence of BH binaries. For instance, recent observations have detected an abundant number of binary systems that contain stellar-mass BHs in the central parsec of the Galactic Centre, where the supermassive BH, Sagittarius A* resides~\cite{2018Natur.556...70H}. This finding is in agreement with the current models of galactic stellar dynamics, which also predicts a population of isolated BHs and of BBHs in this central galactic region. Thus, BBHs are expected to be common astrophysical systems.

Theoretical and phenomenological properties of BBHs have  been studied for a long time - see, $e.g.$~\cite{1987thyg.book..330T,Schutz:1989yw,Kulkarni:1993fr,Sigurdsson:1993zrm,Colpi:2003wb,2016PhRvL.116f1102A,Belczynski:2016ieo}. A particularly interesting feature is their strong lensing effect. Like stationary isolated BHs, dynamical BBHs bend light in their proximity creating deformed images, or even multiple images of background bright objects. Moreover, these dynamical sources cast  {\it shadows} - regions in the local sky lacking radiation, associated with null geodesics that, when propagated backwards in time are absorbed by the BHs (see~\cite{Perlick:2004tq,Cunha:2018acu} for reviews). Solving for the lensing effects, including their shadows, of general-relativistic BBHs is, however, more challenging than for isolated cases. The spacetime geometry created by astrophysical binaries is {\it dynamical} and not known analytically. Thus, the lensing effects/shadows are typically resolved to high accuracy via performing ray tracing on top of dynamical fully non-linear numerical simulations. Specific features of the shadows of BBHs have been identified in these numerical studies. For instance, in dynamical BBHs there are two prominent visible shadows, each associated with one of the two BHs, with narrow secondary `eyebrow' shadows close to the outside of each primary shadow. Such eyebrows also occur in static double BH configurations~\cite{Yumoto:2012kz,Nitta:2011in,Shipley:2016omi,Cunha:2018gql}. In this static binary systems one typically has axial symmetry, with the lensing images, that include aligned eyebrows and main shadows, manifesting this symmetry.  In dynamical BBHs, by contrast, both the intrinsic spin of each BH and the orbital spin of the binary are responsible for frame-dragging, producing a shift of the eyebrow's position in the direction opposite to the spin, as shown in \cite{2015CQGra..32f5002B}.  

Motivated by the recent BBHs discoveries, in this paper we report on a computationally simpler method to reproduce, as a proxy, what an observer in the vicinity of a BBH would see, due to the strong lensing of light induced by the dynamical binary. In particular, this conceptually simple method is able to reproduce the leading effects of the orbital angular momentum of the binary. 

\begin{figure}[H]
\begin{center}
\includegraphics[width=0.4\textwidth]{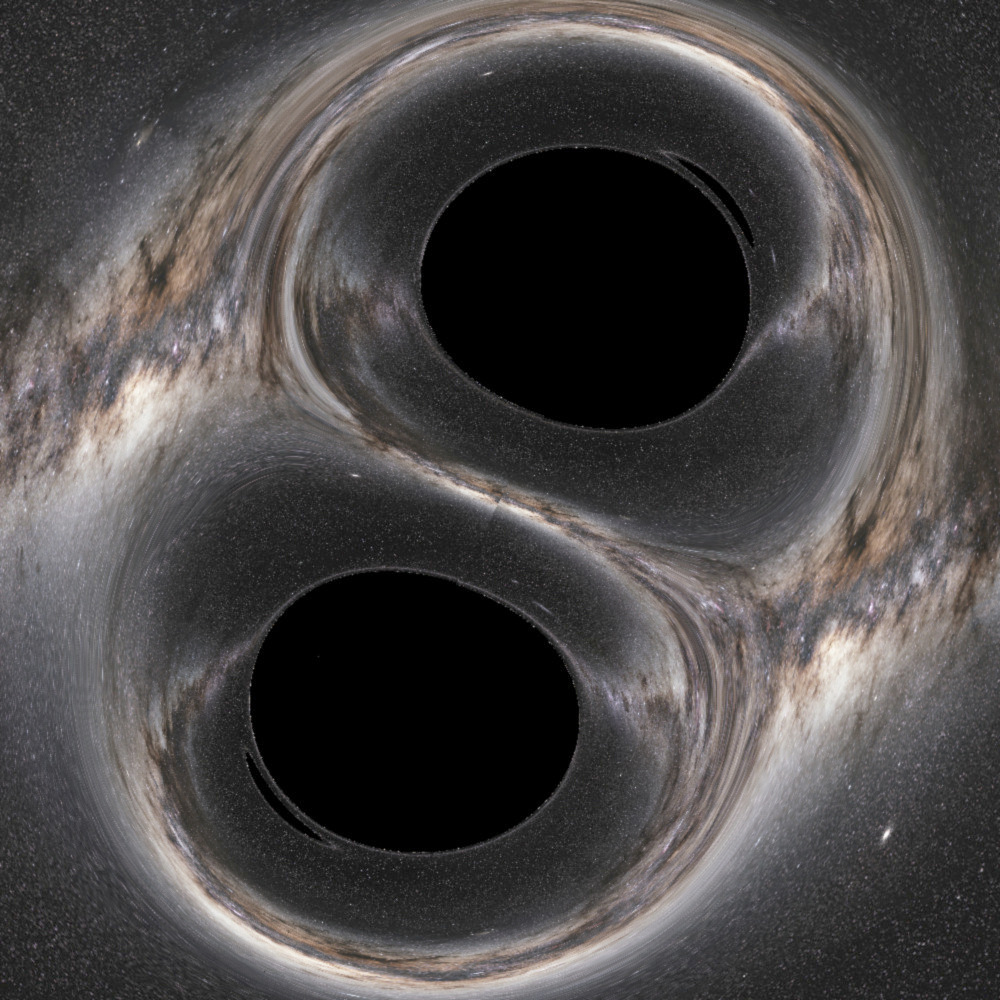}
\caption{\small 
Shadows and lensing in a quasi-static binary BH, using ESO's Milky Way sky~\cite{ESO-sky} as background. The separation between the (equal mass) BHs is $z_o=3M$, where M is the mass of each component, and the counter-clockwise rotation is $\omega M=0.02$ (see Section 2). Supplementary movies can be found in \cite{webpage}.} 
\label{stars}
\end{center}
\end{figure}

The method presented herein is based on a {\it quasi}-static approach to resolve the photon orbits for BBHs. The strategy is to locally compute null geodesics on top of an exact \textit{static or stationary} BBH background, such as the double Schwarzschild (a.k.a. Bach-Weyl) geometry~\cite{Bach:1922}, and periodically adjust them by small rigid rotational corrections along an axial vector field that does not coincide with the axi-symmetry of the exact solution, thus mimicking the orbital spin of the BHs. These corrections along the photon positions will be discrete  rotations, with the frame of the two BHs fixed. This procedure provides a proxy to computing the paths of light rays that meet an observer in the vicinity of a truly dynamical binary. A snapshot of such a {\it quasi}-static evolution of the geodesics on a static double-Schwarzschild BH solution \cite{Israel:1966rt}, using an image of the Milky Way as background, is depicted in Fig.~\ref{stars}. Supplementary movies for the shadows and lensing due to this quasi-static BBHs can be found in \cite{webpage}. As we shall illustrate below the leading characteristic features of the shadows of the full dynamical BBHs are replicated by this procedure. 

To study the effects of the \textit{intrinsic} (rather than orbital) angular momenta of the BHs in the BBHs system we also compute the shadows of stationary (non dynamical) spinning BBHs solutions of general relativity. We shall use the double-Kerr BH solutions \cite{Herdeiro:2008kq,Manko:2017avt,Cabrera-Munguia:2017dol,Costa:2009wj,Manko:2013iva} that are known exactly. These are asymptotically flat metrics that represent two Kerr BHs with a conical singularity between them (in the representation we use). Similar conical defects are found in the double-Schwarzschild BH solutions. In the latter case, it was observed in~\cite{Cunha:2018gql} that thanks to the underlying cylindrical symmetry, of both the geometry and the spatial part of the fundamental photon orbits, the conical singularity has essentially no observable effect on the shadows. Since a cylindrical symmetry is also present in the double Kerr solution, we also expect no observable effect in the shadows due to the conical singularities. This is confirmed by computing the null geodesics in these backgrounds: we are able to produce images of the shadows of the co-rotating (even) and counter-rotating (odd) exact stationary BBHs configurations. Our resulting images show complex and in some cases self-similar structure across different angular scales. Among the stationary BBHs there is a set of extremal, maximally spinning solutions. The extremal configurations have finite size (event horizon area) and zero temperature. While we present a few images of the shadows of stationary BBHs spinning near extremality, images of the first representations of the exactly extremal BBHs will be presented in a forthcoming publication, where we shall make contact with the recent analysis of the near horizon geometry of these BH binaries~\cite{Ciafre:2018jpe}.

In what follows we describe, in Section 2, the shadows of \textit{quasi}-static BBHs, to address the effect of orbital angular momentum on the lensing. We focus on explaining the method we developed to trace light rays and present our results. In Section 3, we turn to the effect of the intrinsic spin on the lensing, considering the double-Kerr BBHs where we do not consider any kind of orbital spin. In section \ref{section:even}, we compute the shadows of the stationary double-Kerr BH \cite{Manko:2017avt,Cabrera-Munguia:2017dol} with co-rotating (even) spins. And in section \ref{section:odd} we find and analyze the shadows of the stationary counter-rotating (odd) double-Kerr BH \cite{Costa:2009wj,Manko:2013iva}. These more analytical approaches that we introduce (compared to ray tracing on numerical simulations) will hopefully enable a better understanding of the shadows of astrophysical BBHs.

\section{\textit{Quasi}-static Binary Black Holes}
\label{section:quasi}

The double-Schwarzschild BH is a {\it static} solution of the vacuum Einstein's equations. Starting from it, however, we can construct a rotation proxy that mimics the leading effects of a fully dynamical BH binary, in what concerns lensing effects. In this section we will describe such a proxy, focusing on signatures at the level of the shadows.

The rotation of the binary will be assumed to be {\it adiabatic-like}, $i.e.$ the BHs will move rather slowly when compared with the light ray travel time for a typical photon reaching the observer. Under this approximation, photons will locally follow null geodesics in the double-Schwarzschild (static) background, with the trajectory periodically suffering small corrections due to the rotation of the BHs. These corrections will simply be discrete (active) rotations of the photon position along their path, with the frame of the two BHs fixed, with such a procedure being straightforward to implement numerically. At the end of the trajectory the photon position is rotated back into the observer's frame. This system will be dubbed a \textit{quasi}-static BH binary.

The static BH binary (the double Schwarzschild solution) will be described in Weyl coordinates $x^\alpha=(t,\rho,\varphi,z)$ - see~\cite{Cunha:2018gql} for the details of the solution. Consider then a map 
\begin{eqnarray}
\Omega: &&\mathcal{M} \to \mathcal{M} \\  
&& x^{\alpha} \to  x^{\alpha'}=\Omega^{\alpha'}(x^{\alpha})\ ,
\end{eqnarray}
where $x^{\alpha'}=(t',\rho',\varphi',z')$, that takes each point of our manifold $\mathcal{M}$ to another point in $\mathcal{M}$. In order to naively mimic a Cartesian rotation, the map $\Omega$  is defined as follows (in Weyl coordinates):

\[\
 \begin{cases}
     t'=t\\
     \rho'=\sqrt{x'^2+y'^2}\\
     \varphi'=\left\{
    \begin{array}{l}
    \displaystyle{ \textrm{asin}\frac{y'}{\rho'} \quad \textrm{if}\quad(x'\geqslant 0)}\\
    \displaystyle{ \pi-\textrm{asin}\frac{y'}{\rho'}\,\, \textrm{if}\quad(x'<0)} \end{array} \right. \\
     z'= z',
   \end{cases} \ ,
    \left(
    \begin{array}{l}
    x' \\
    y' \\
    z' \\
    \end{array}
    \right) =
    \left(
    \begin{array}{ccc}
    1 & 0 & 0 \\
    0 & \cos \omega\delta t  & \sin\omega\delta t \\
    0 &-  \sin\omega\delta t &  \cos \omega\delta t  \\
    \end{array}
    \right) 
     \left(
    \begin{array}{l}
    x \\
    y \\
    z \\
    \end{array}
    \right) \ ,
   \]
and $x+iy=\rho e^{i\varphi}$.
Hence, after a time interval $\delta t$, the photon position is corrected by changing its initial location $P=(t,\rho,\varphi,z)$ to a new point $P'=(t',\rho',\varphi',z')$ under the map $\Omega$. We remark that $\Omega$ is well defined even when $(\omega\delta t)\gg 1$, although this will not be usually the case during the numerical integration of the trajectory.

The next issue is how the photon's 4-momentum should be modified. The vector $p=p^\mu\partial_\mu$ at point $P$ can be projected via $\Omega$ into the push-forward vector $(\Omega^*p)$ at $P'$~\cite{Carroll:1997ar,Wald:1984rg}:
\begin{align*}
(\Omega^*p)=p^\mu\partial_\mu\Omega^{\alpha'} \partial_{\alpha'}= p^t\partial_{t'} + (p^i\partial_i\Omega^{a'}) \partial_{a'},
\end{align*}
where $i\in\{\rho,\varphi,z\}$ and $a'\in\{\rho',\varphi',z'\}$. However, restrictions have to be imposed to $(\Omega^*p)$, in order for it to represent the 4-momentum at $P'$. We impose the new momentum $\widetilde{p}$ should satisfy the following two requirements:
\begin{itemize}
\item The photon's local energy $\mathcal{E}$ is the same for a static observer in $P$ and $P'$, $i.e.$ $\mathcal{E}=\sqrt{-g_{tt}}\,p^t=\sqrt{-g'_{tt}}\,\widetilde{p}^{t'}$~\cite{Cunha:2016bjh}. This is reasonable because the physical rotation is performed by the BHs, and an observer in $P$ can be identified with one in $P'$.
\item The norm of $\widetilde{p}$ vanishes, $i.e.$ $\widetilde{p}^{\alpha'}\,\widetilde{p}_{\alpha'}=0$.
\end{itemize}
It follows that the new momentum $\widetilde{p}$ at $P'$ is then defined as
\[\widetilde{p}= \left(\sqrt{\frac{g_{tt}}{g'_{tt}}}\,p^t\right) \partial_{t'} + \gamma\left(p^i\partial_i\Omega^{a'}\right)\partial_{a'},\]
where the (positive) factor $\gamma$ enforces the vanishing of the norm. We further remark that this procedure modifies the values of the photon's energy $E=-p_t$ and axial angular momentum $L=p_\varphi$ with respect to infinity, which otherwise would be Killing constants of motion. This implies that a photon can in principle escape the system with more (or less) energy than it started with. We stress that this operation does not amount (generically) to a simple coordinate transformation.

Although the angular frequency $\omega$ of the BH binary is a free parameter that was introduced in the model, a physically reasonable value of $\omega$ can be estimated from the Keplerian orbital frequency:
\[\omega\sim \left(\Delta z+1\right)^{-3/2}\,M^{-1},\]
where $M$ is the ADM mass, and $\Delta z$ is the proper distance between the two BHs. The latter can be computed with a complete elliptic integral of the second kind (see~\cite{PhysRevD.80.104036}):
\[\Delta z =(2z_o+1)\left(1-\frac{1}{4z_o^2}\right)E\left(\frac{2z_o-1}{2z_o+1}\right).\]
The parameter $z_o$ in the previous expression parametrises the BH distance and is the same that was used in~\cite{Cunha:2018gql}.

\begin{figure}[H]
\begin{center}
\includegraphics[width=0.35\textwidth]{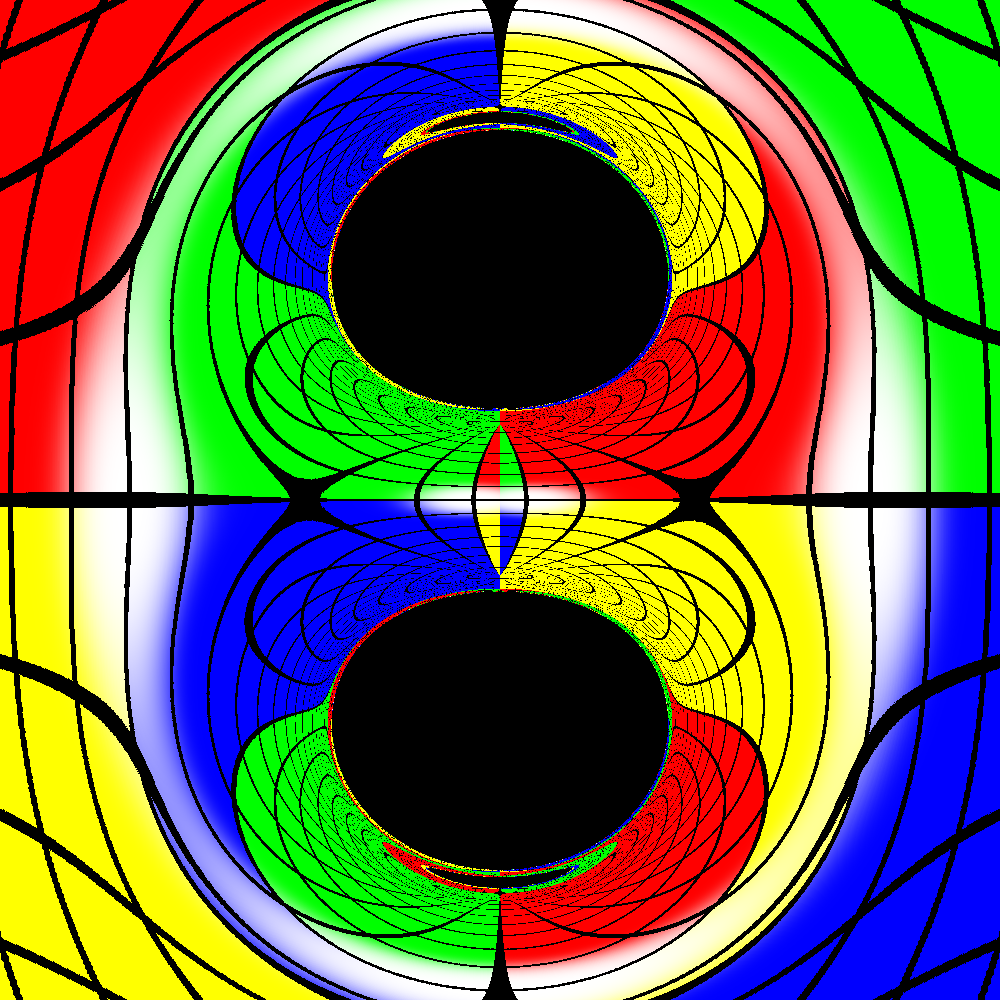}
\ \ \ \includegraphics[width=0.35\textwidth]{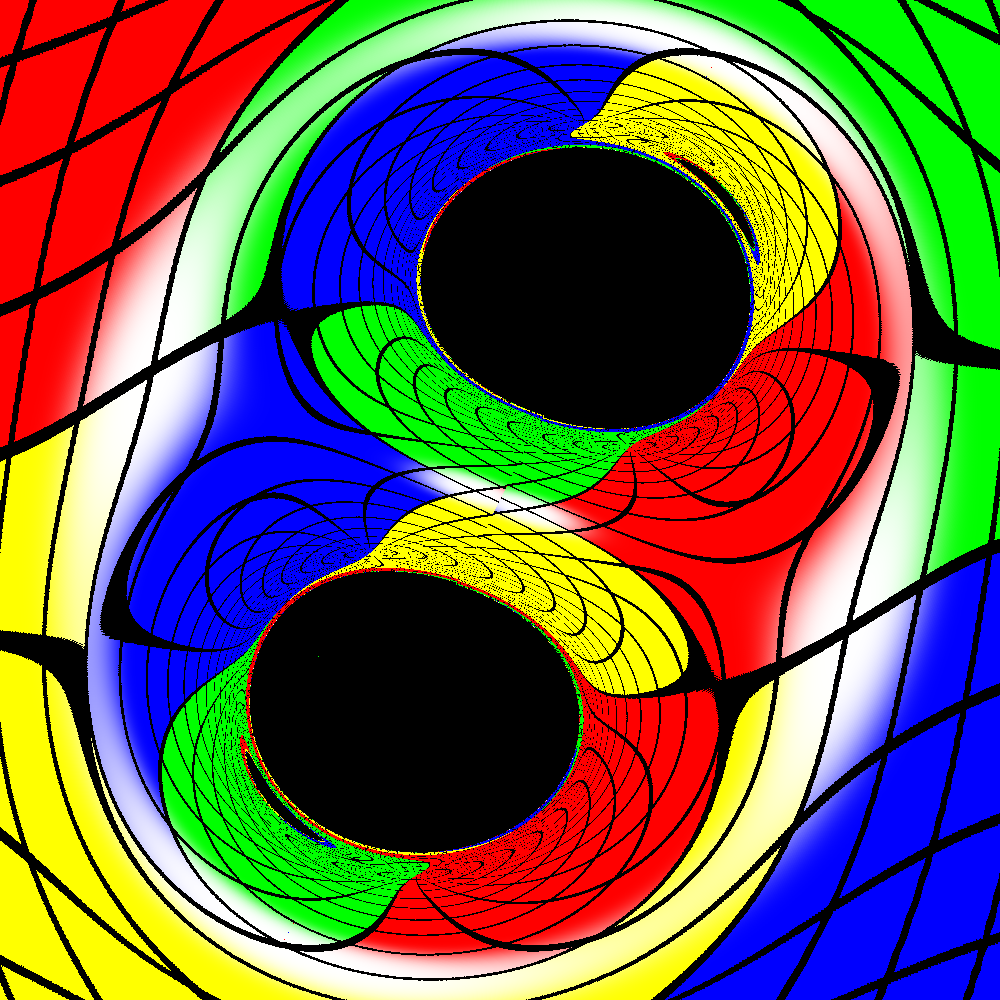}
\includegraphics[width=0.35\textwidth]{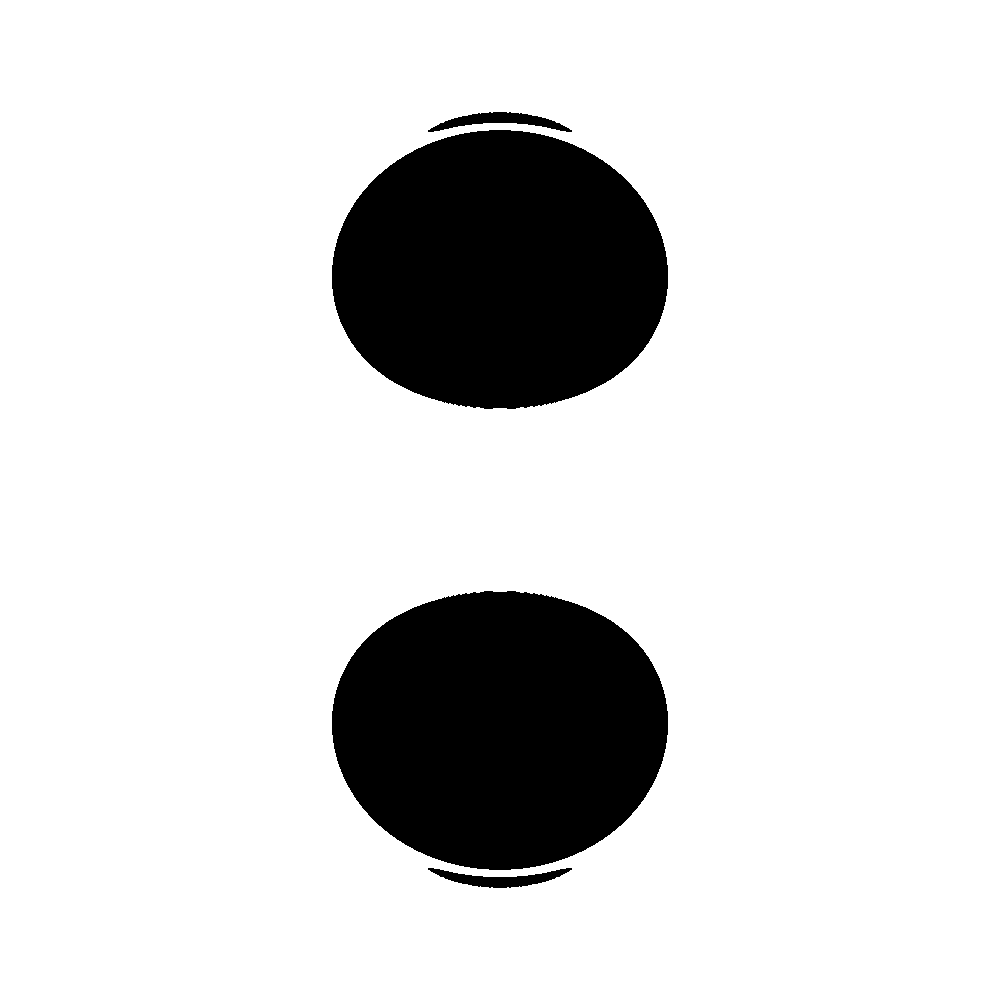} \ \ \ 
\includegraphics[width=0.35\textwidth]{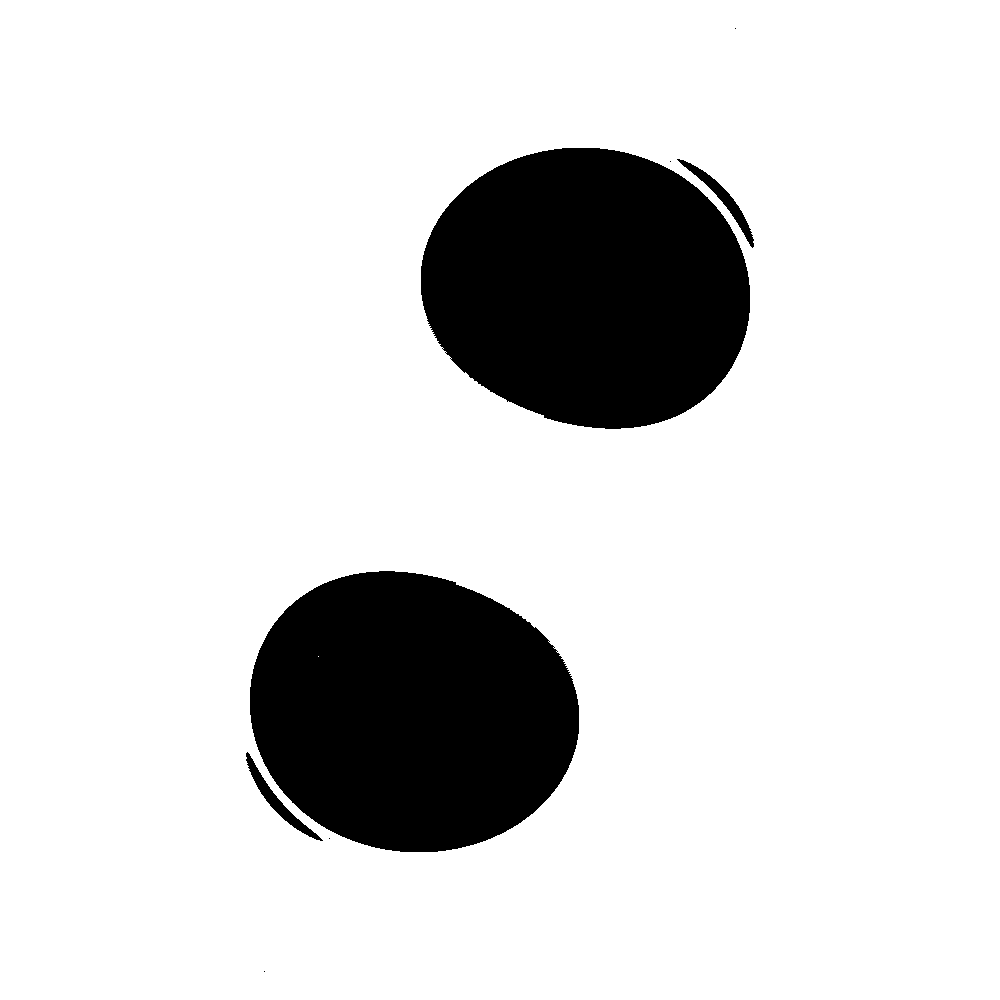}
\caption{\small \textit{Top row:} Lensing of a static (left) quasi-static (right) BH binary with $z_o=3M$ and $\omega M=\{0\,,\,0.02\}$. \textit{Bottom row:} Shadows of the previous images. The observer sits along the axis of orbital rotation (the $x$-axis). The BHs rotate counter-clockwise in the image for positive $\omega$.}
\label{rotation}
\end{center}
\end{figure}

Implementing the approach we have just described to the double Schwarzschild solution, using the same setup as used in~\cite{Cunha:2018gql} (see also~\cite{Cunha:2015yba,Cunha:2016bjh,Cunha:2017eoe}),  we obtain the lensing and shadows displayed in~Fig.~\ref{rotation}.  The first column of this figure  displays the lensing and shadows of a static double Schwarzschild BH with $z_o=3M$, already discussed in~\cite{Cunha:2018gql}. The second column displays the corresponding quasi-static binary with $\omega=0.02 M^{-1}$, with the BHs rotating \textit{counter-clockwise} in the image (see movie in~\cite{webpage}). Observe that the shadows are twisted \textit{clockwise} in the image with respect to the static case. This can be interpreted as follows. The observation image was taken at coordinate time $t=0$; at this time the binary had the same vertical orientation as in the static case. Since light takes a finite amount of time to get to the observer, the shadows are actually recording the BH positions at a past time ($t<0$), when the BHs were rotated clockwise with respect to $t=0$. The shadow eyebrows are a second order lensing effect, related to a time even further into the past, thus presenting an additional clockwise rotation with respect to the main shadows. For illustration purposes we have included Fig.~\ref{stars} with the same lensing and shadows of the rightmost image of Fig.~\ref{rotation}, but replacing the colored background with an image of the Milky Way (see movie in~\cite{webpage}).

\begin{figure}[H]
\begin{center}
\includegraphics[width=0.4\textwidth]{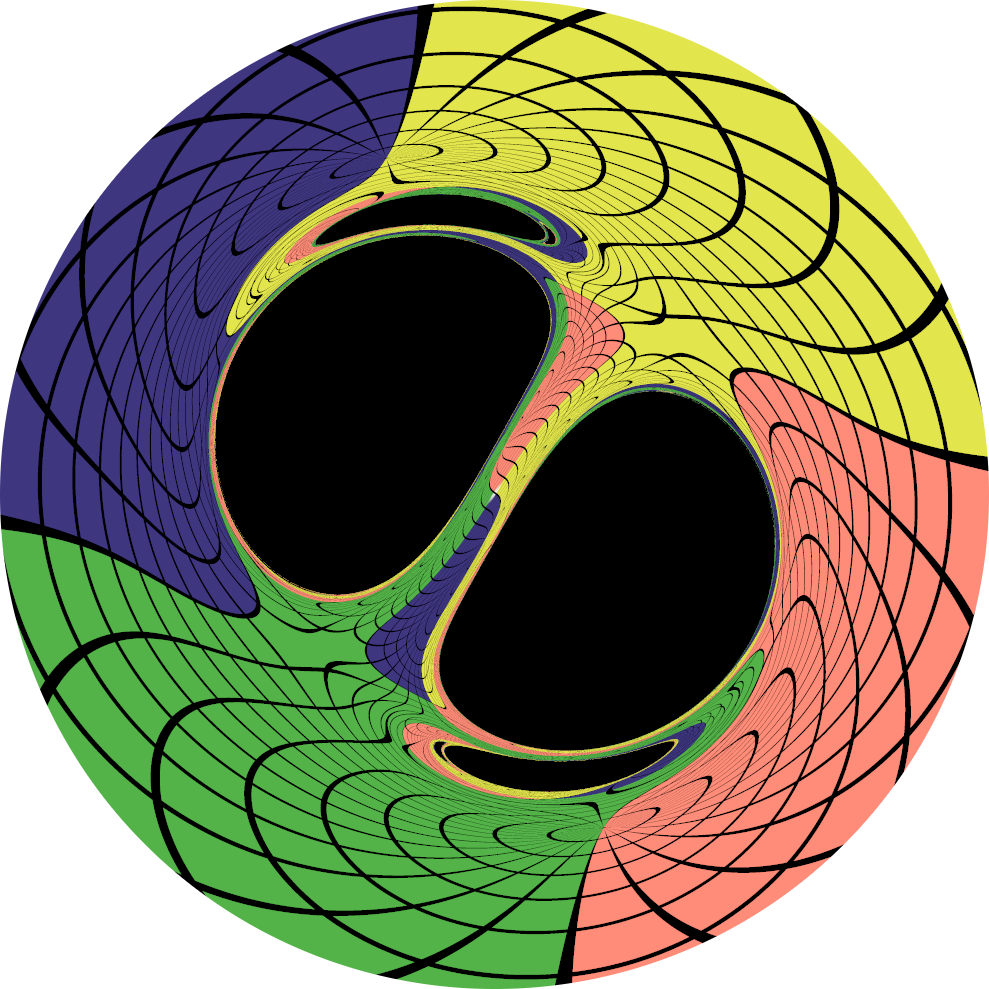} \ \ \ 
\includegraphics[width=0.4\textwidth]{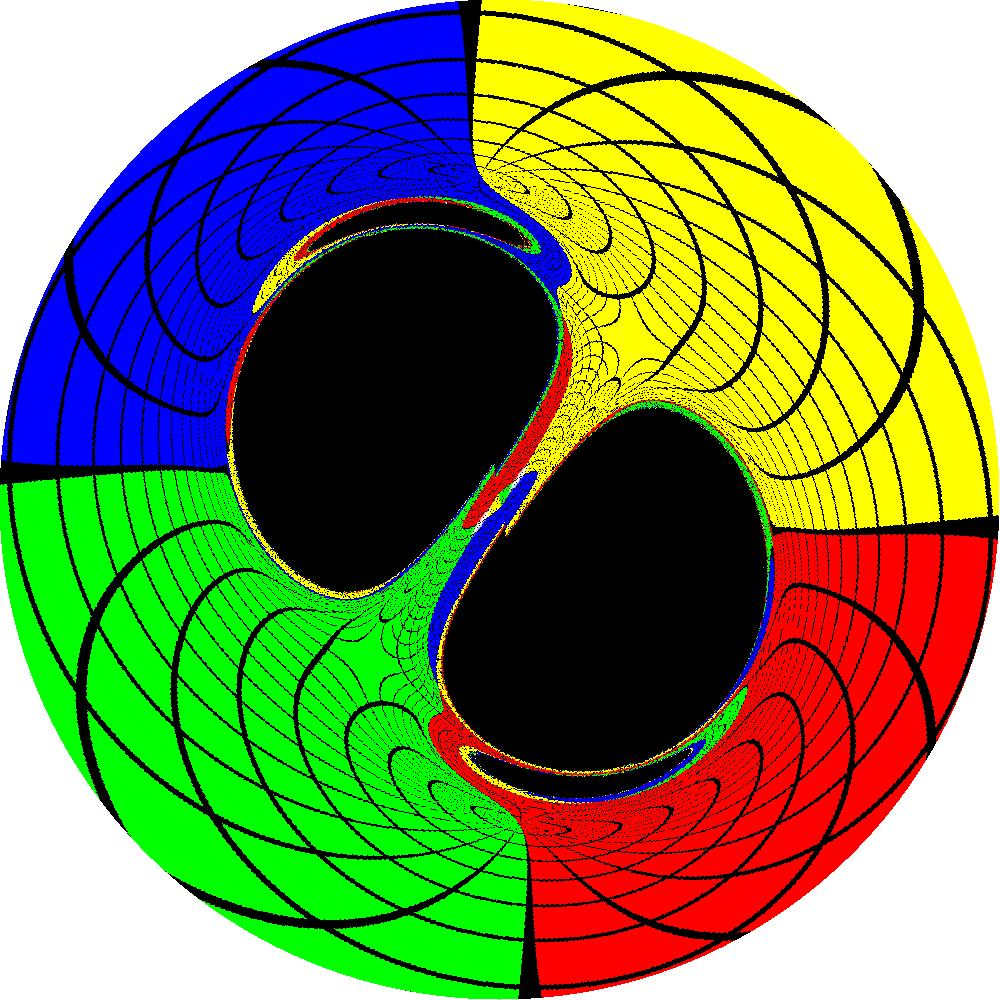}
\caption{\small {\it Left:} Shadows and lensing of a fully dynamical binary of equal-mass BHs with no
spin (adapted from~\cite{2015CQGra..32f5002B}). {\it Right:} Quasi-static BH binary with $z_o=1.5M$ and $\omega=0.06\,M^{-1}$.}
\label{bohn}
\end{center}
\end{figure}

To assess the accuracy of the method described above as a proxy to the lensing in a dynamical BBH, we perform, in Fig.~\ref{bohn}  a lensing comparison of a fully dynamical binary in~\cite{2015CQGra..32f5002B} close to merger with a similar quasi-static binary. Despite clear differences concerning specific details of the lensing,  the overall \textit{qualitative} resemblance between both cases at the level of the shadow structure is uncanny. Still, in order to have a more \textit{quantitative} comparison between both images in Fig.~\ref{bohn}, we define two parameters $\chi,\psi$. The first parameter, $\chi$, is the ratio between the shadow area\footnote{The shadow area corresponds to a solid angle in the observer's sky.} of the main shadows  and the one of the associated eyebrows; one obtains $\chi\simeq \{15,20\}$, respectively for the left (right) image of Fig.~\ref{bohn}.
The second parameter, $\psi$, is an angle that parametrises the eyebrows' angular displacement with respect to the main shadows. By first computing the average position of all points within each shadow element, one can draw two straight lines connecting similar average position points, $e.g.$ eyebrow to eyebrow and main shadow to main shadow. It is then possible to define $\psi$ as the angle formed between these two lines. We obtain $\psi\simeq\{42^{\circ}, 33^{\circ}\}$ respectively for the left (right) image of Fig.~\ref{bohn}. We remark that $\psi=0$ for the static BH binary, by symmetry (see left image of Fig.~\ref{rotation}).

Although the values $\chi,\psi$ are not exactly the same for both images in Fig.~\ref{bohn}, the quasi-static binary is here displayed mainly as a proof of concept. In particular, the values of $\{z_o,\omega\}$ of the quasi-static binary are quite \textit{ad-hoc}, leaving some room for optimization. Moreover, note that we have chosen a binary BBH close to merger, in which case the adiabatic approximation of the quasi-static binary is starting to break down, as the BHs change their positions on a time scale comparable to the light ray's travel time towards the observer. In addition, unusual effects at the level of the shadow start to be noticeable, in particular a non-smooth edge ($i.e.$ a cusp) due to the combination of the conical singularity and rotation.\\

\section{Stationary Binary Black Holes}
The previous section illustrated the mimicked effect of the orbital angular momentum of a dynamical binary in the lensing of light, by using the double Schwarzschild solution in a quasi-static approximation. Generically, however, dynamical binaries also have intrinsic BH spin. We now show that the lensing effect of the intrinsic spin is taking into account by considering a \textit{stationary} binary, rather than static, described by the double Kerr solution. We shall be interested in the particular cases of the Kerr solution describing two equal mass BHs with either equal (even case) or opposite (odd case) spins. In both cases the double Kerr solution has a conical singularity in between the BHs and is described by the line element, in Weyl coordinates:
\begin{equation}
ds^2= -f(dt-\omega\,d\varphi)^2 + \frac{e^{2\gamma}}{f}\left(d\rho^2 +dz^2\right) +\frac{\rho^2}{f}d\varphi^2,
\end{equation}
where the metric functions $f,\gamma,\omega$ only depend on the coordinates $\rho,z$.

\subsection{The even case}
\label{section:even}
\begin{figure}[H]
\begin{center}
\includegraphics[width=0.22\textwidth]{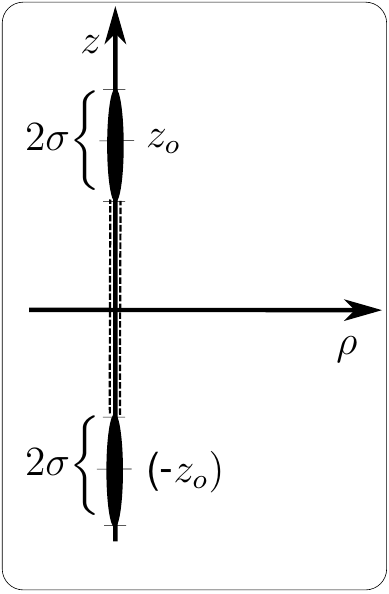}
\caption{\small Schematic representation of the equal double-Kerr BH system with identical BHs. The solid black rods along the $z$-direction represent each a BH while the dashed line in between these rods correspond to the conical singularity. The quantity $\sigma$ is proportional to the horizon temperature~\cite{Cabrera-Munguia:2017dol}.} 
\label{setup}
\end{center}
\end{figure}

For two equal mass and equal spin BHs, the metric functions are defined as~\cite{Manko:2017avt,Cabrera-Munguia:2017dol}:
\[f=\frac{A\bar{A}-B\bar{B}}{(A+B)(\bar{A}+\bar{B})},\quad e^{2\gamma}=\frac{A\bar{A}-B\bar{B}}{K_o^2R_{11}R_{01}R_{10}R_{00}}\quad \omega=4a-\frac{2\textrm{Im}\left\{(\bar{A}+\bar{B})G\right\}}{A\bar{A}-B\bar{B}},\]
where the overbar denotes complex conjugation and
\[A=4z_o^2(R_{11}-R_{01})(R_{10}-R_{00})-4\sigma^2\left(R_{11}-R_{10}\right)\left(R_{01}-R_{00}\right),\]
\[B=8z_o\sigma\bigg[(z_o+\sigma)(R_{01}-R_{10})-(z_o-\sigma)(R_{11}-R_{00})\bigg],\]
\vspace{0.2cm}
\[G=-zB + 8z_o\sigma\bigg[z_o(R_{01}R_{00}-R_{11}R_{10}) +\sigma(R_{11}R_{01}-R_{10}R_{00}) -(z_o^2-\sigma^2)(R_{11}-R_{01}-R_{10}+R_{00})\bigg],\]
\vspace{0.2cm}
\[R_{jk}(\rho,z)=\frac{-2(\epsilon\sigma+\kappa z_o)+2id}{1+4(\kappa z_o+ia)(\epsilon\sigma+ia)}\sqrt{\rho^2+(z+\kappa z_o+\epsilon\sigma)^2},\qquad \epsilon=2j-1,\quad \kappa=2k-1,\]
\vspace{0.2cm}
\[\sigma=\sqrt{\frac{1}{4}-a^2 + d^2\left(4z_o^2-1+4a^2\right)^{-1}},\qquad K_o=16\sigma^2\left\{\frac{(2z_o^2+z_o+2a^2)^2-a^2}{(z_o+1/2)^2+a^2}\right\},\]
\vspace{0.2cm}
\[d=\frac{a(4z_o^2-1+4a^2)}{(4z_o^2+2z_o+4a^2)},\]
with quantities normalized to the ADM mass $M$ of the solution. This solution has two free parameters, $z_o$ and $a$, with $z_o$ denoting the coordinate position of each BH in the $z$-axis (see Fig.~\ref{setup}), whereas $a$ is a spin parameter related to ADM axial angular momentum $J=2a-d$.

\begin{figure}[ht!]
\begin{center}
\includegraphics[width=0.5\textwidth]{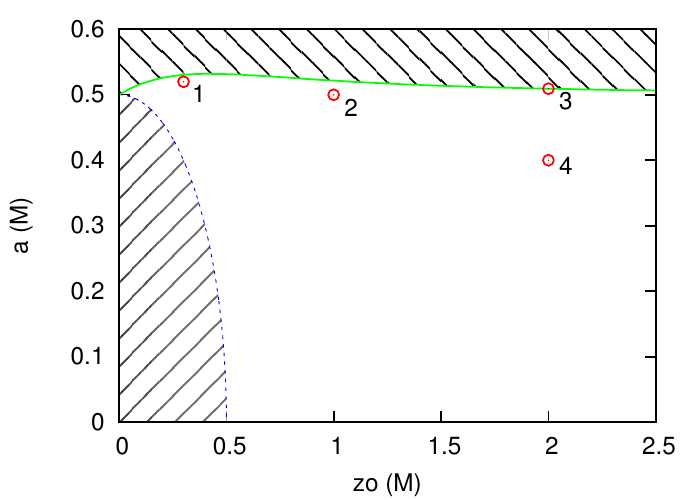}
\caption{\small Parameter space ($z_o,a$) of the double-Kerr solution with identical co-rotating BHs. The shaded regions are considered unphysical, with the dashed (solid) line representing the limit II (III). The shadows of the configurations 1 $\to$ 4 are displayed in Fig.~\ref{shadows-2Kerr}.} 
\label{domain}
\end{center}
\end{figure}

The physical domain of the parameter space $\{z_o,a\}$ obeys the condition $z_o\geqslant \sigma\geqslant 0$, with $\sigma$ and all metric functions real. The domain with $a\geqslant 0$ has the following limits (see Fig.~\ref{domain}):
\begin{enumerate}
\item[I.] Double-Schwarzschild solution ($a=0\implies J=0$), with $z_o\geqslant 1/2$;
\item[II.] Single Kerr BH, given by $\sigma=z_o$; this leads to $a^2 + z_o^2=1/4$ (blue dashed line in Fig.~\ref{domain});
\item[III.] Extremal limit, provided by $\sigma=0\implies$ vanishing temperature  (green solid line in Fig.~\ref{domain});
\item[IV.] Two isolated Kerr BHs with $z_o\to \infty$. 
\end{enumerate}
We remark that there is an additional independent region which also satisfies $z_o\geqslant \sigma\geqslant 0$ but for which the metric can have Closed-Timelike-Curves (CTCs)~\cite{Wald:1984rg}; it is thus discarded as unphysical.

\begin{figure}[H]
\begin{center}
\includegraphics[width=0.25\textwidth]{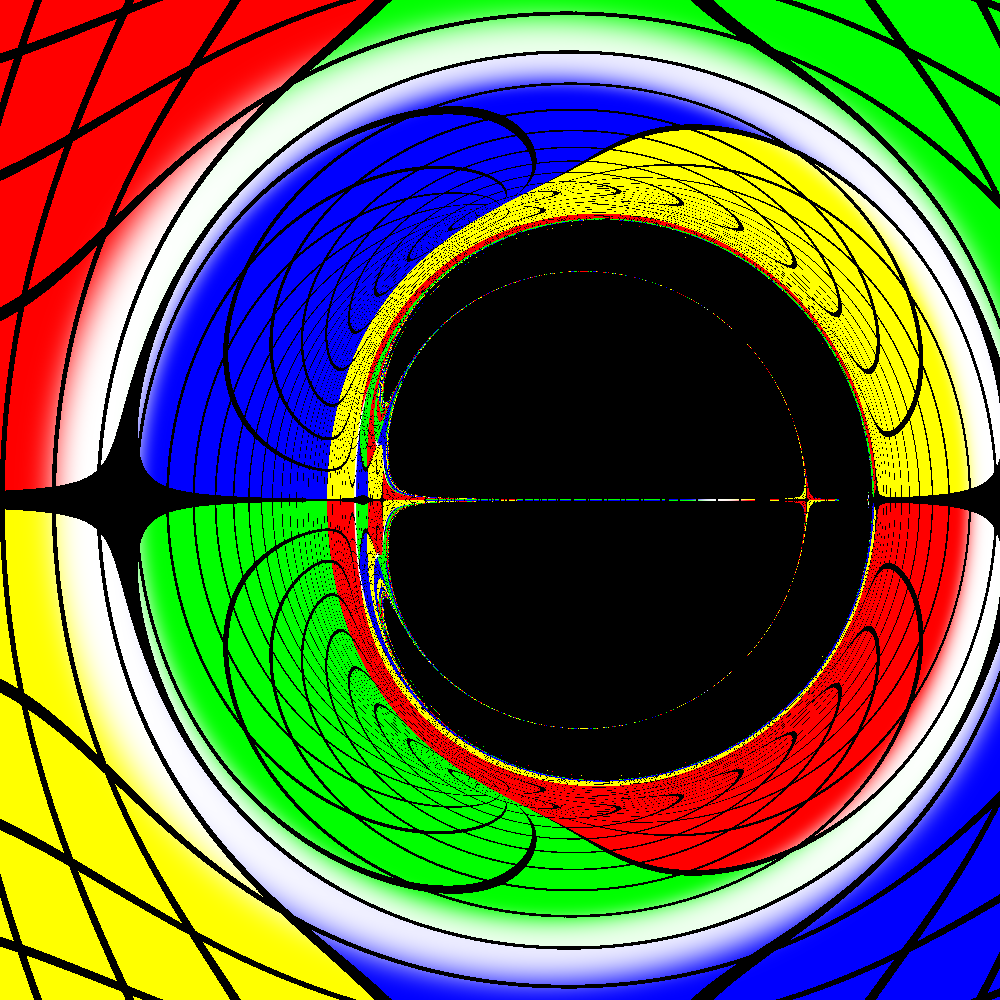}\includegraphics[width=0.25\textwidth]{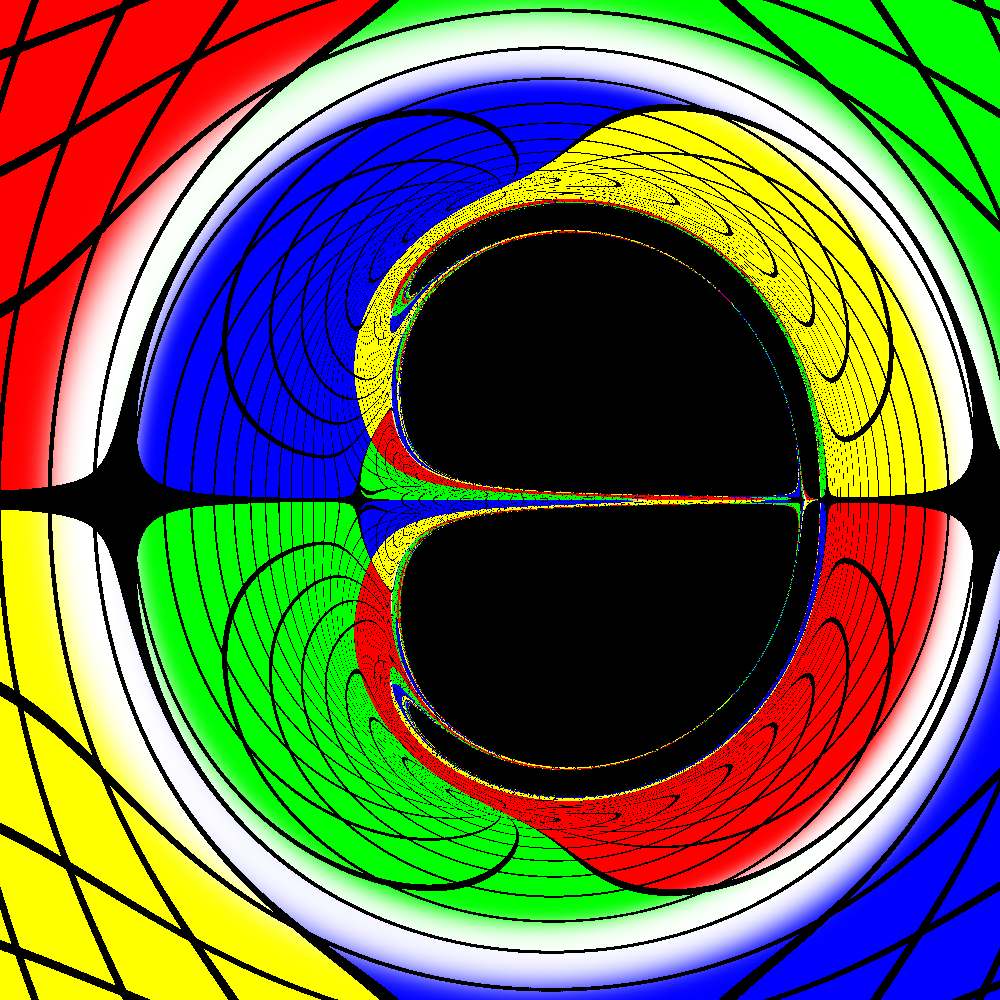}\includegraphics[width=0.25\textwidth]{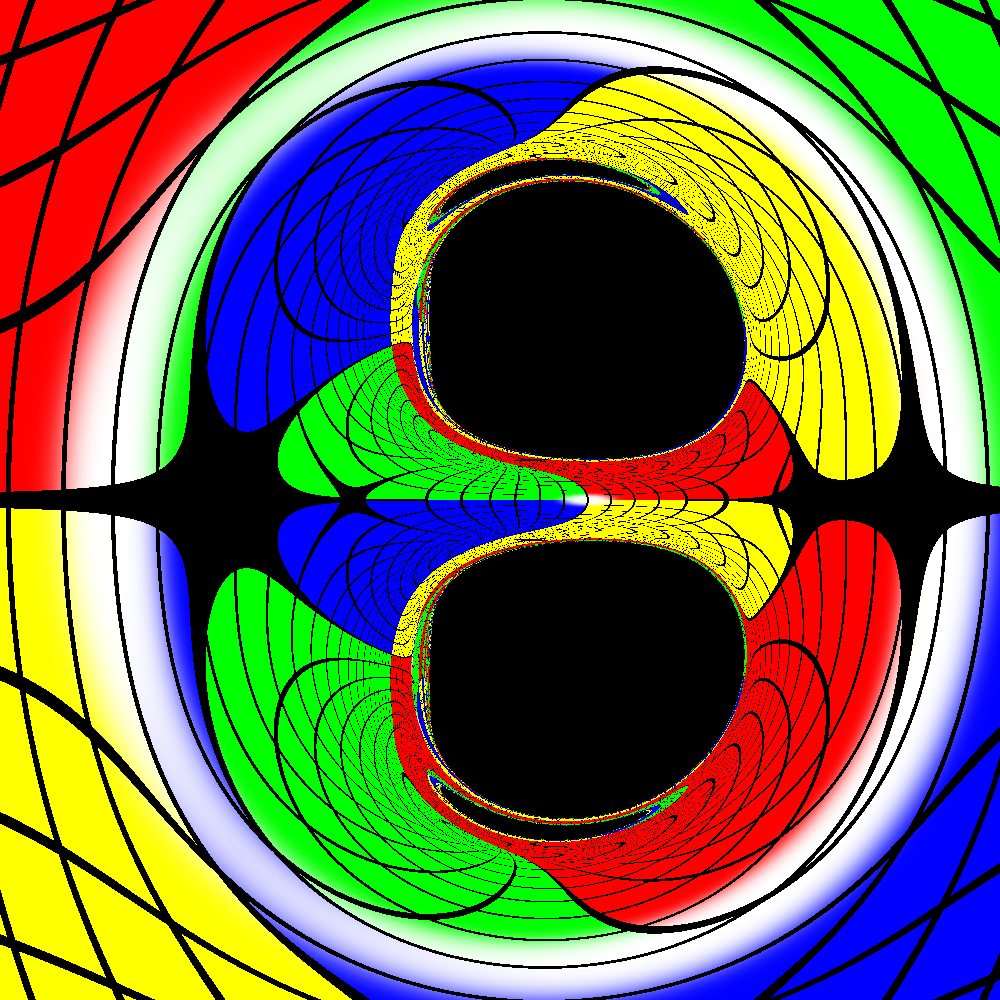}\includegraphics[width=0.25\textwidth]{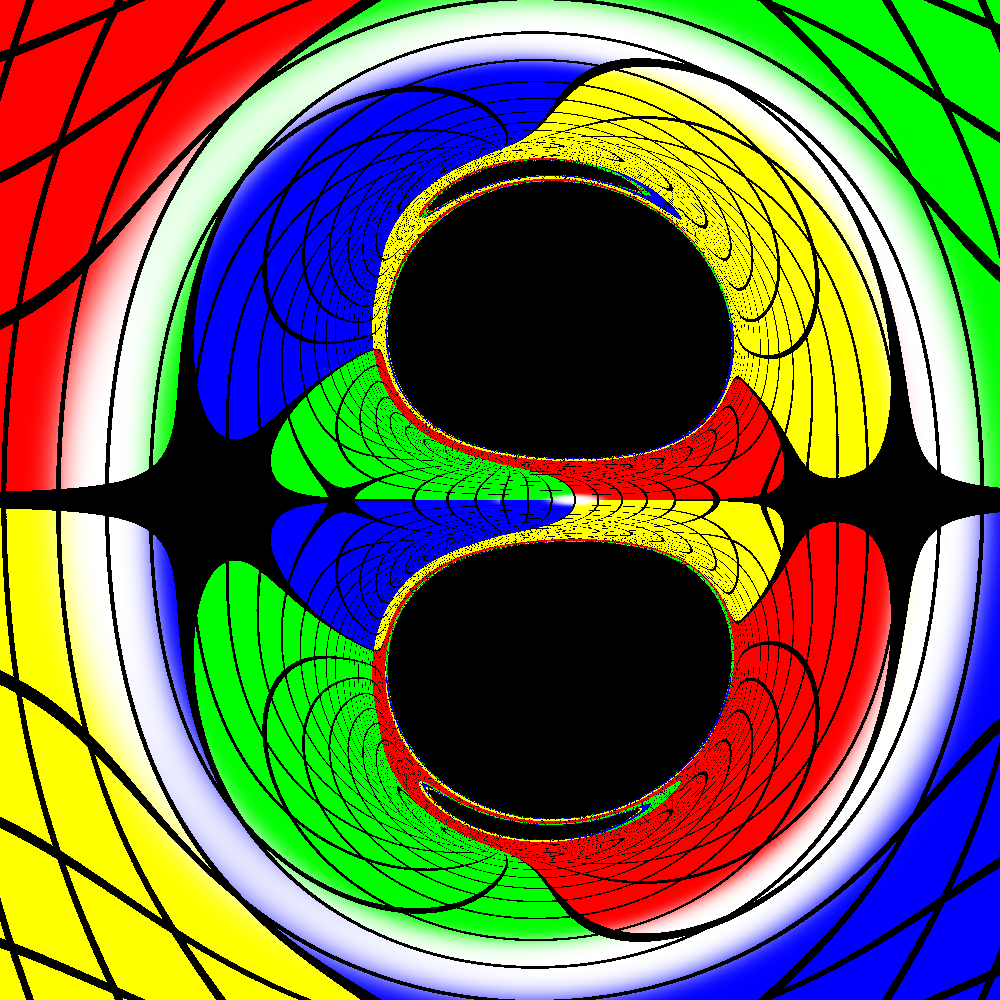}\\
\includegraphics[width=0.25\textwidth]{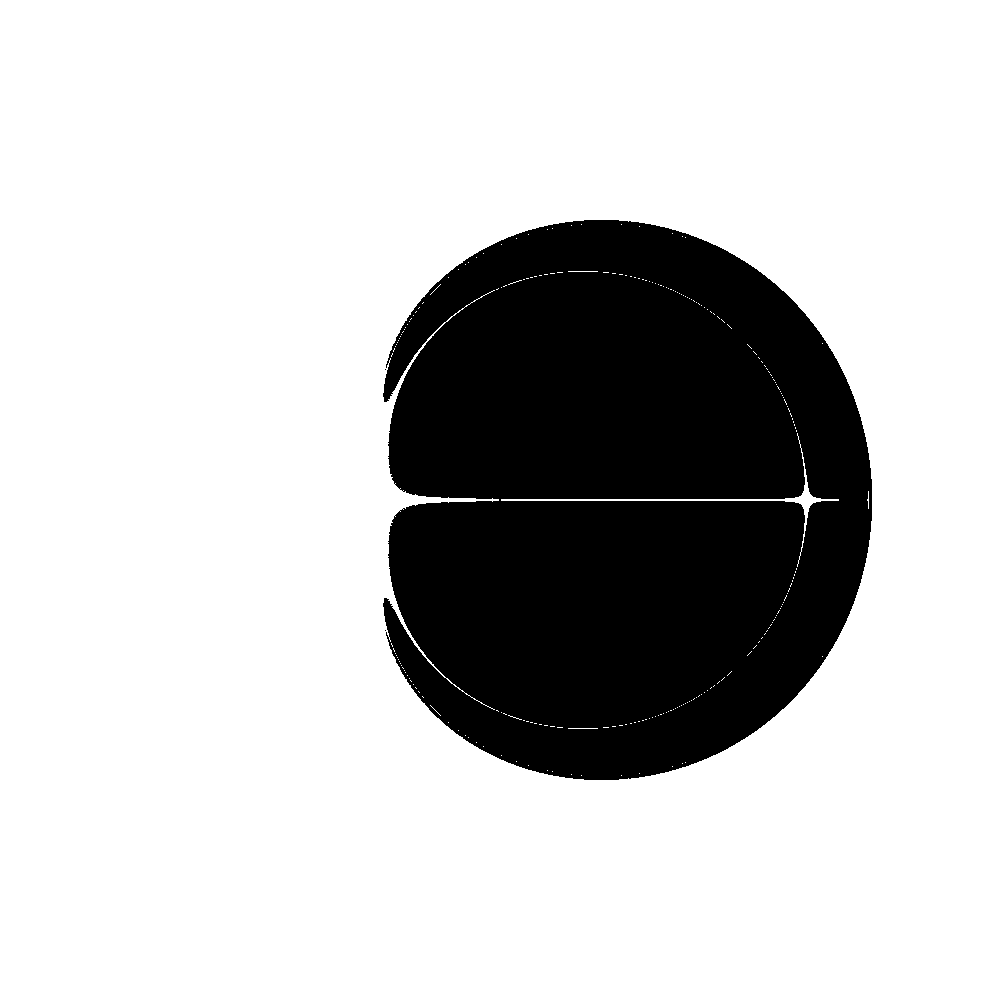}\includegraphics[width=0.25\textwidth]{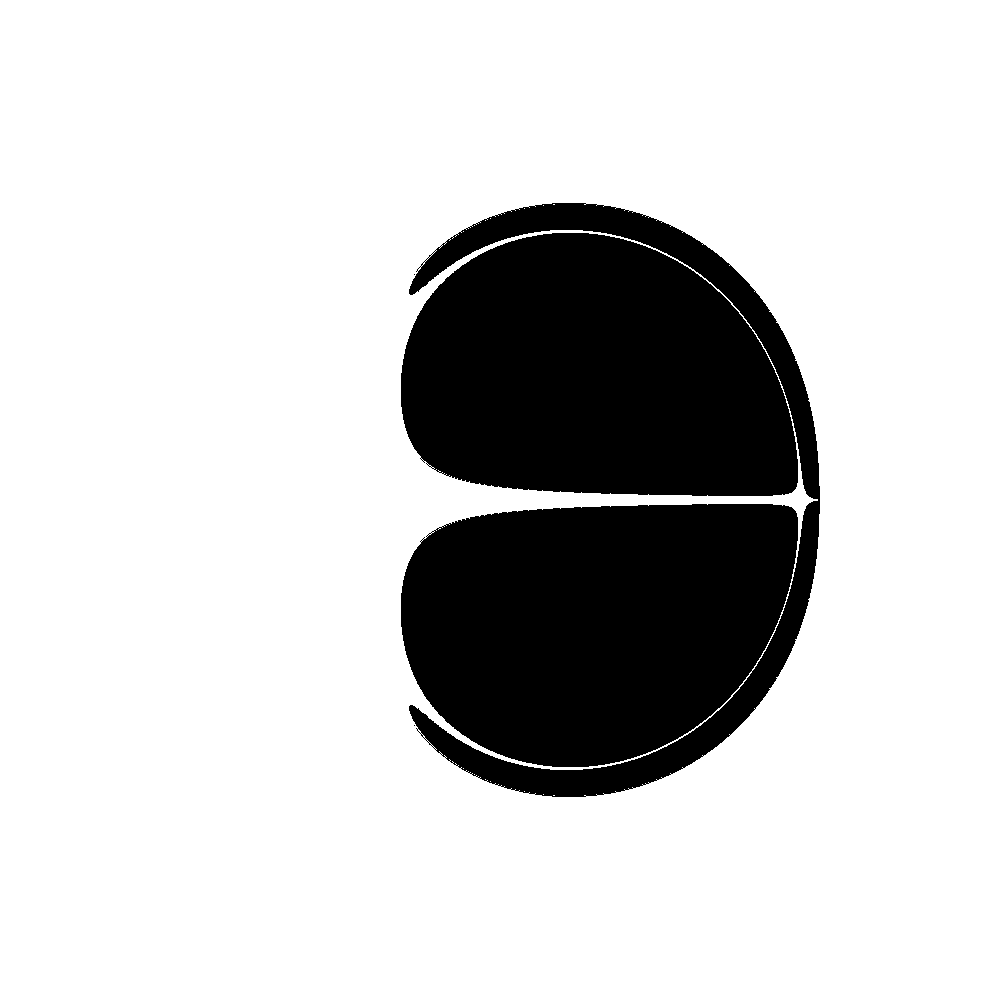}\includegraphics[width=0.25\textwidth]{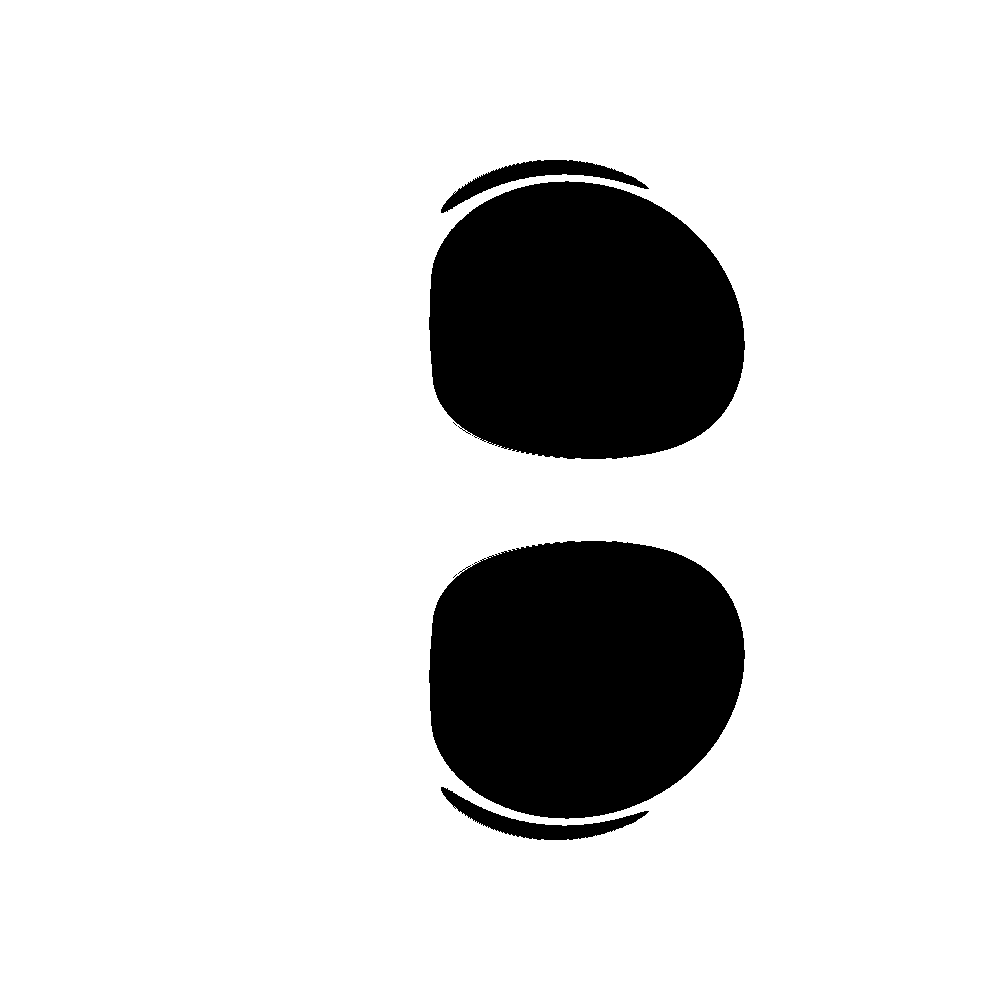}\includegraphics[width=0.25\textwidth]{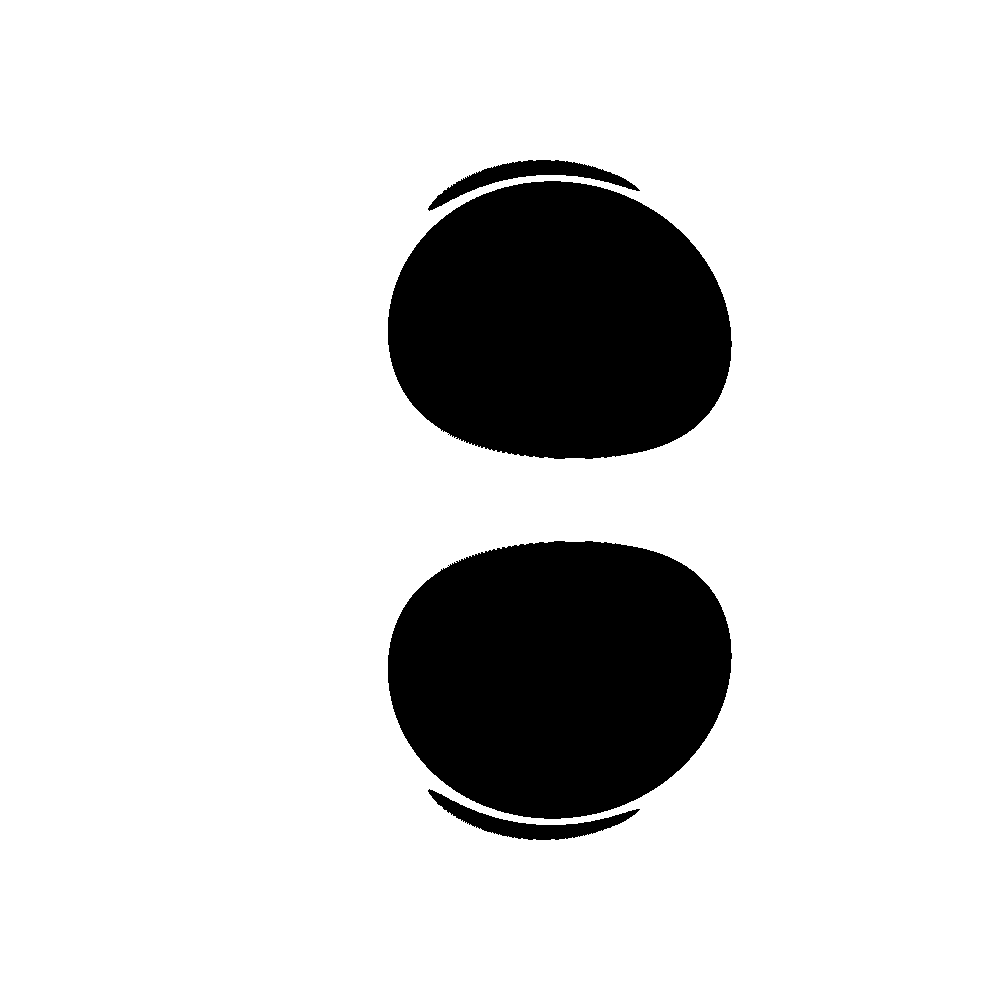}\\
\includegraphics[width=0.25\textwidth]{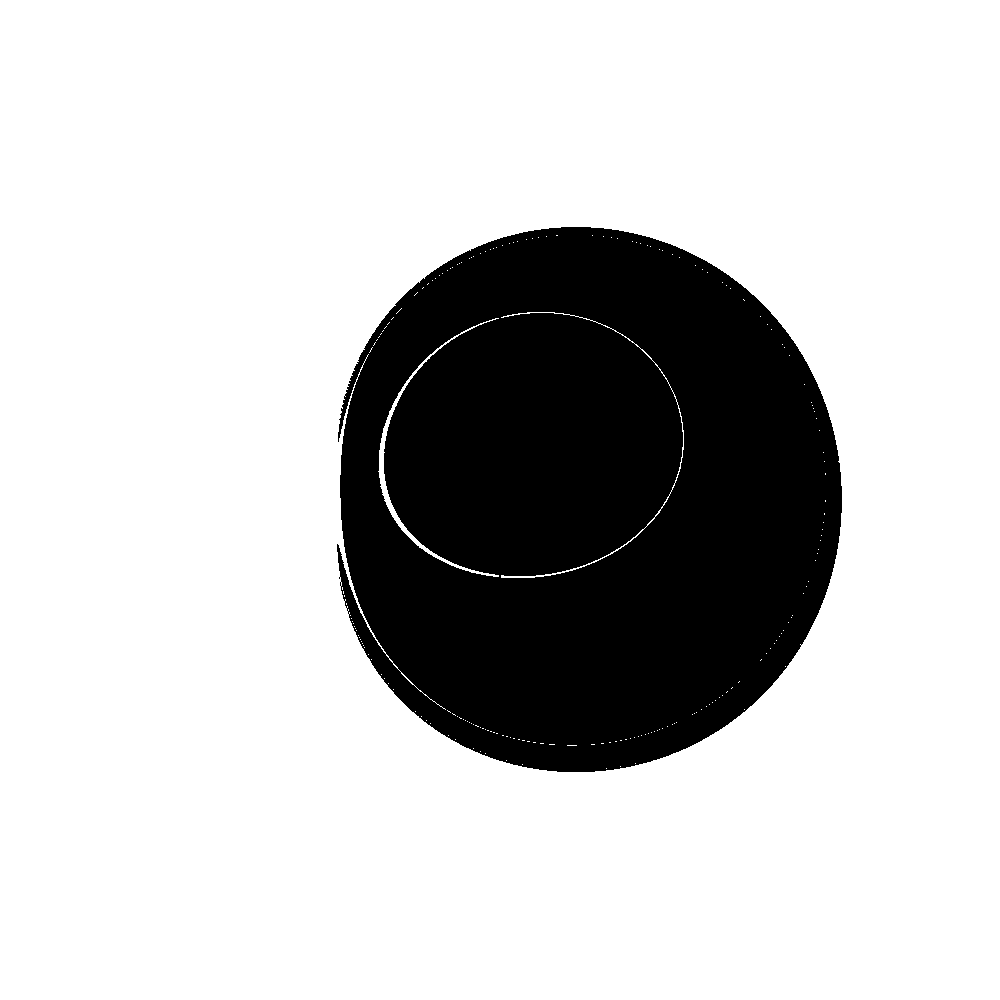}\includegraphics[width=0.25\textwidth]{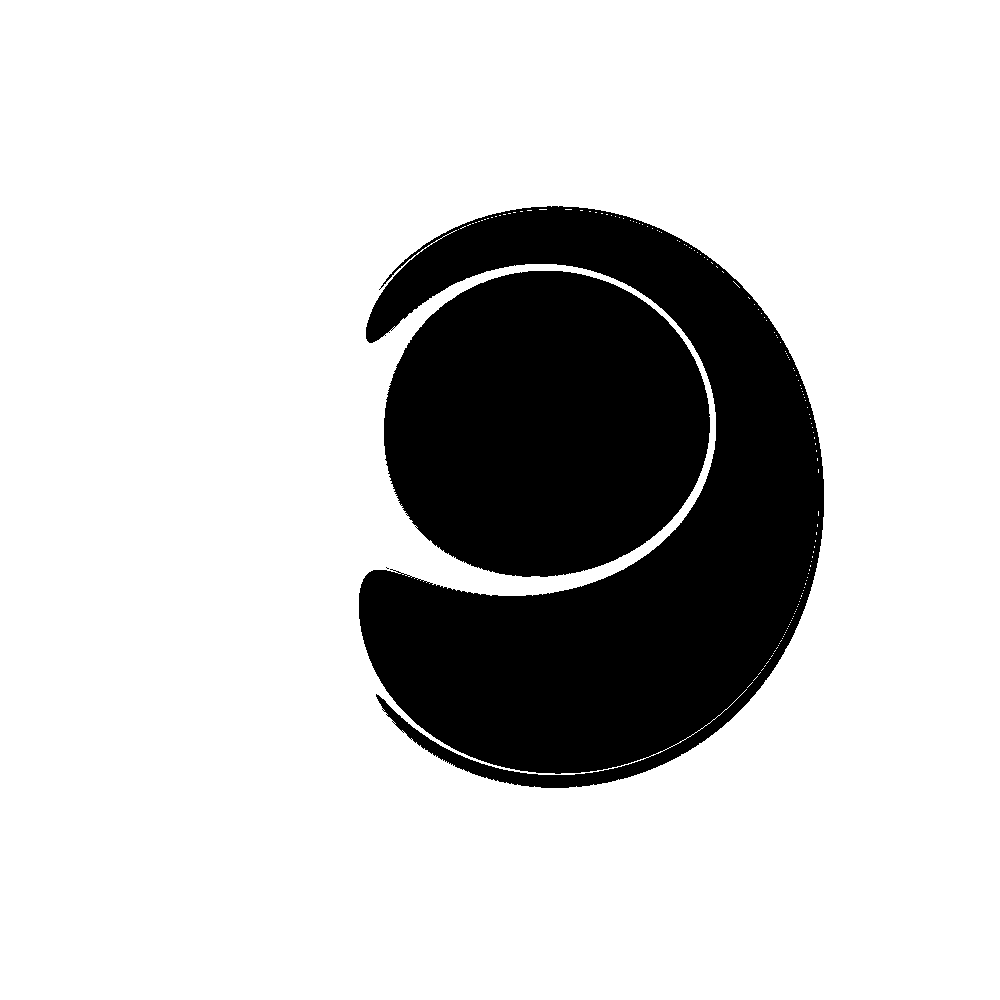}\includegraphics[width=0.25\textwidth]{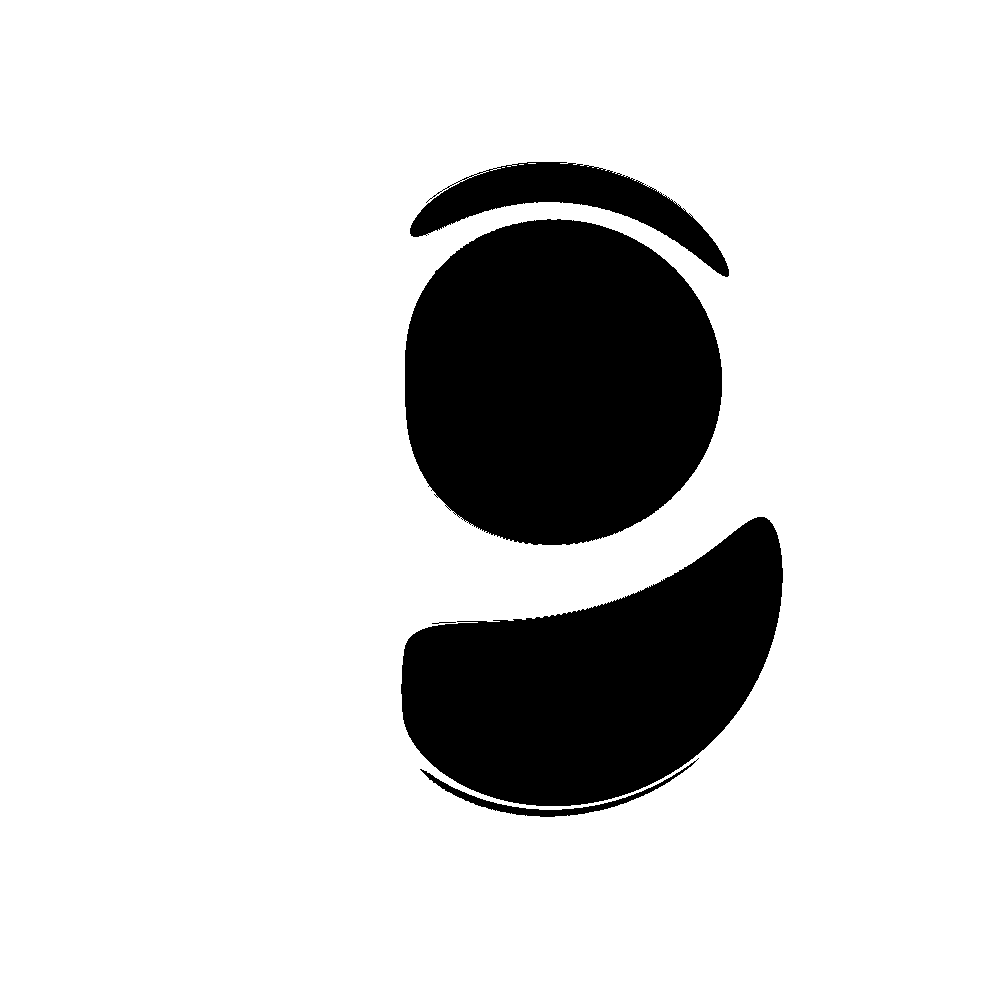}\includegraphics[width=0.25\textwidth]{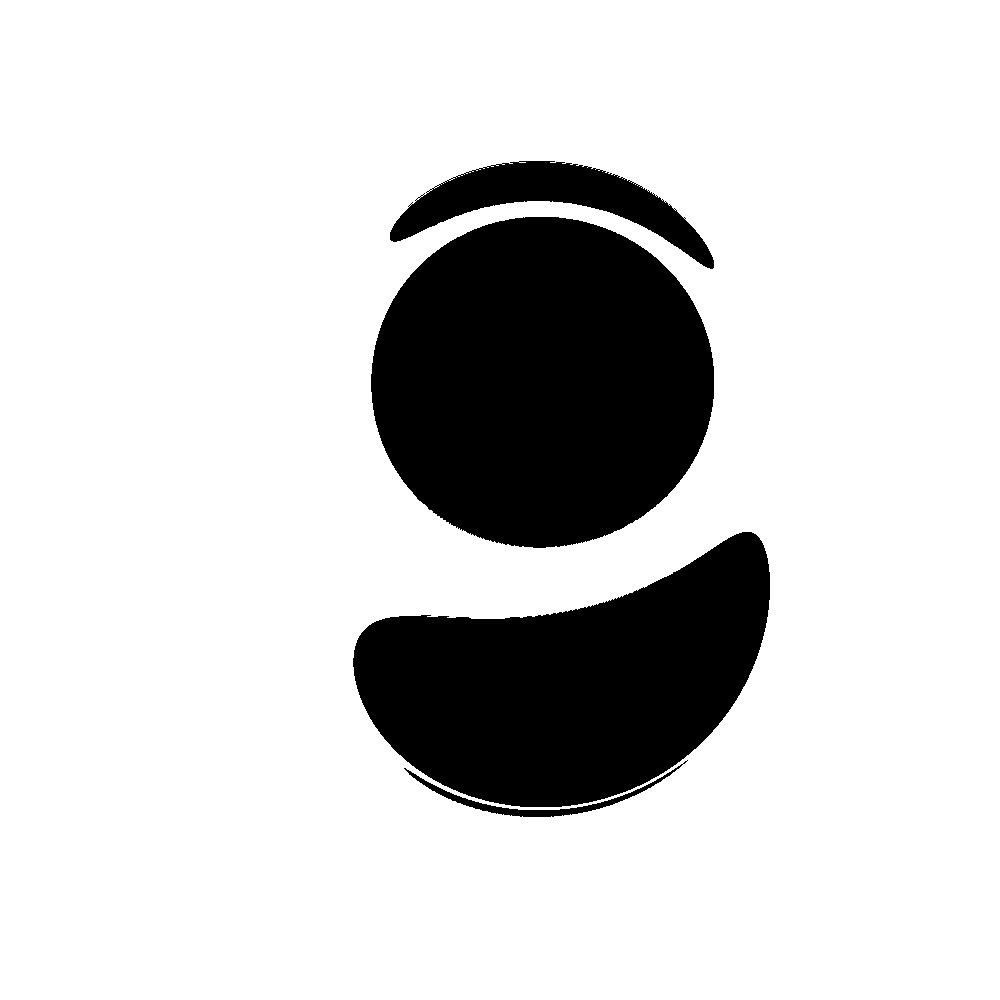}
\caption{\small Lensing of configurations 1 $\to$ 4 of Fig.~\ref{domain} (columns from left to right). The second (third) row displays only the shadows, as observed with $\theta_o=\pi/2$ ($\theta_o=\pi/4$).}
\label{shadows-2Kerr}
\end{center}
\end{figure}

The shadows and lensing of four solutions, marked in Fig.~\ref{domain} with red dots, are displayed in Fig.~\ref{shadows-2Kerr}. There appear to be no strikingly new features in the shadows. In particular the D-like shadow profile, characteristic of a fast spinning (single) Kerr BH, still holds in the double Kerr case (namely solution 3), as one could have naively anticipated. The third row of Fig.~\ref{shadows-2Kerr} also displays observations outside the equatorial plane, with $\theta_o=\textrm{acos}(z/\sqrt{z^2+\rho^2})=\pi/4$.

\subsection{The odd case}
\label{section:odd}
For two equal mass and opposite spin BHs, the metric functions are defined as~\cite{Costa:2009wj,Manko:2013iva,Manko:2008pv}:
\[f=\frac{A\bar{A}-B\bar{B}}{(A+B)(\bar{A}+\bar{B})},\quad e^{2\gamma}=\frac{A\bar{A}-B\bar{B}}{(4z_o\sigma)^4R_{11}R_{01}R_{10}R_{00}}\quad \omega=-\frac{2\textrm{Im}\left\{(\bar{A}+\bar{B})G\right\}}{A\bar{A}-B\bar{B}},\]
where the overbar denotes complex conjugation and
\[A=\sigma^2(R_{11}R_{01}+R_{10}R_{00}) +z_o^2(R_{11}R_{10}+R_{01}R_{00})+\]
\[+(R_{11}R_{00}+R_{01}R_{10})\left(\frac{z_o}{2}+\sigma^2[8z_o^2-1]\right) -4ia\sigma z_o(2z_o-1)(R_{11}R_{00}-R_{01}R_{10}),\]
\vspace{0.2cm}
\[B=4\sigma^2z_o^2(R_{11}+R_{01}+R_{10}+R_{00})-\sigma z_o\bigg(1 +2ia[2z_o-1]\bigg)(R_{11}-R_{01}-R_{10}+R_{00}),\]
\vspace{0.2cm}
\begin{align*}
G=&-zB + 2\sigma^2z_o(R_{10}R_{00}-R_{11}R_{01})+2\sigma z_o^2(R_{01}R_{00}-R_{11}R_{10}) +\\
&+z_o\sigma(z_o+\sigma)(R_{11}-R_{00})\bigg(4z_o\sigma-1-2ia[2z_o-1]\bigg) +\\&+ z_o\sigma(z_o-\sigma)(R_{01}-R_{10})\bigg(4z_o\sigma+1+2ia[2z_o-1]\bigg),
\end{align*}
\vspace{0.2cm}
\[R_{jk}(\rho,z)=\sqrt{\rho^2+(z+\kappa z_o+\epsilon\sigma)^2},\qquad \epsilon=2j-1,\quad \kappa=2k-1,\]
\vspace{0.2cm}
\[\sigma=\sqrt{\frac{1}{4}- a^2\left(\frac{2z_o-1}{2z_o+1}\right)},\]
with quantities normalized to the ADM mass $M$ of the solution. Again, this solution has two free parameters, $z_o$ and $a$, with $z_o$ denoting the coordinate position of each BH in the $z$-axis (see Fig.~\ref{setup}), whereas $a$ is a spin parameter proportional to the (Komar) angular momentum of the lower BH $J^-=a/2$. We further remark that the total ADM angular momentum vanishes since the upper BH has $J^+=-a/2$.\\

\begin{figure}[ht!]
\begin{center}
\includegraphics[width=0.5\textwidth]{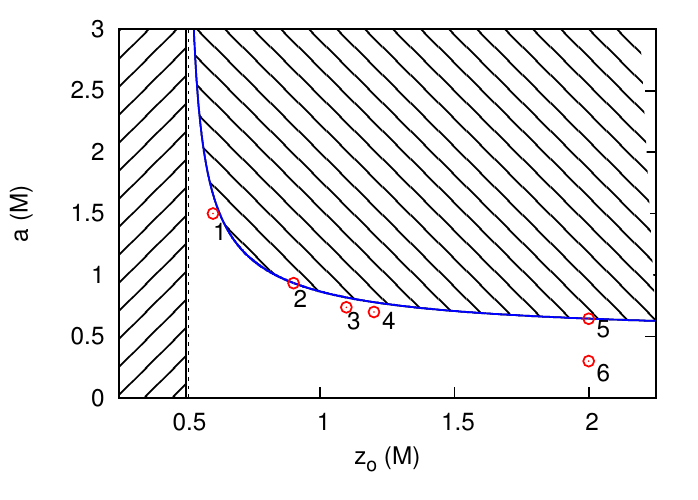}
\caption{\small Parameter space ($z_o,a$) of the double-Kerr (odd) solution with counter-rotating BHs. The shaded regions are considered unphysical. The shadows of the configurations 1 $\to$ 6 are displayed in Fig.~\ref{shadows-2Kerr-odd}.} 
\label{domain-2Kerr-odd}
\end{center}
\end{figure}

Again, the physical domain of the parameter space $\{z_o,a\}$ obeys the condition $z_o\geqslant \sigma\geqslant 0$, with $\sigma$ and all metric functions real. The domain with $a\geqslant 0$ has the following limits (see Fig.~\ref{domain-2Kerr-odd}):
\begin{enumerate}
\item[I.] Double-Schwarzschild solution ($a=0\implies J^{\pm}=0$), with $z_o\geqslant 1/2$;
\item[II.] Single BH, given by $\sigma=z_o=1/2$; (vertical dotted line in Fig.~\ref{domain-2Kerr-odd});
\item[III.] Extremal limit, provided by $\sigma=0\implies a=\pm \frac{1}{2}\sqrt{{(2z_o+1)}/{(2z_o-1)}}$ (blue line in Fig.~\ref{domain-2Kerr-odd});
\item[IV.] Two isolated Kerr BHs with $z_o\to \infty$ and opposite rotation. 
\end{enumerate}
The boundary II corresponds to a Schwarzschild BH when $a=0$ and $z_o=1/2$, whereas for $a\neq 0$ and $z_o=1/2$ the horizon is singular~\cite{Costa:2009wj}. Nevertheless, in terms of shadows and gravitational lensing, the boundary II appears to be indistinguishable from the Schwarzschild case.

\begin{figure}[h!]
\begin{center}
\includegraphics[width=0.25\textwidth]{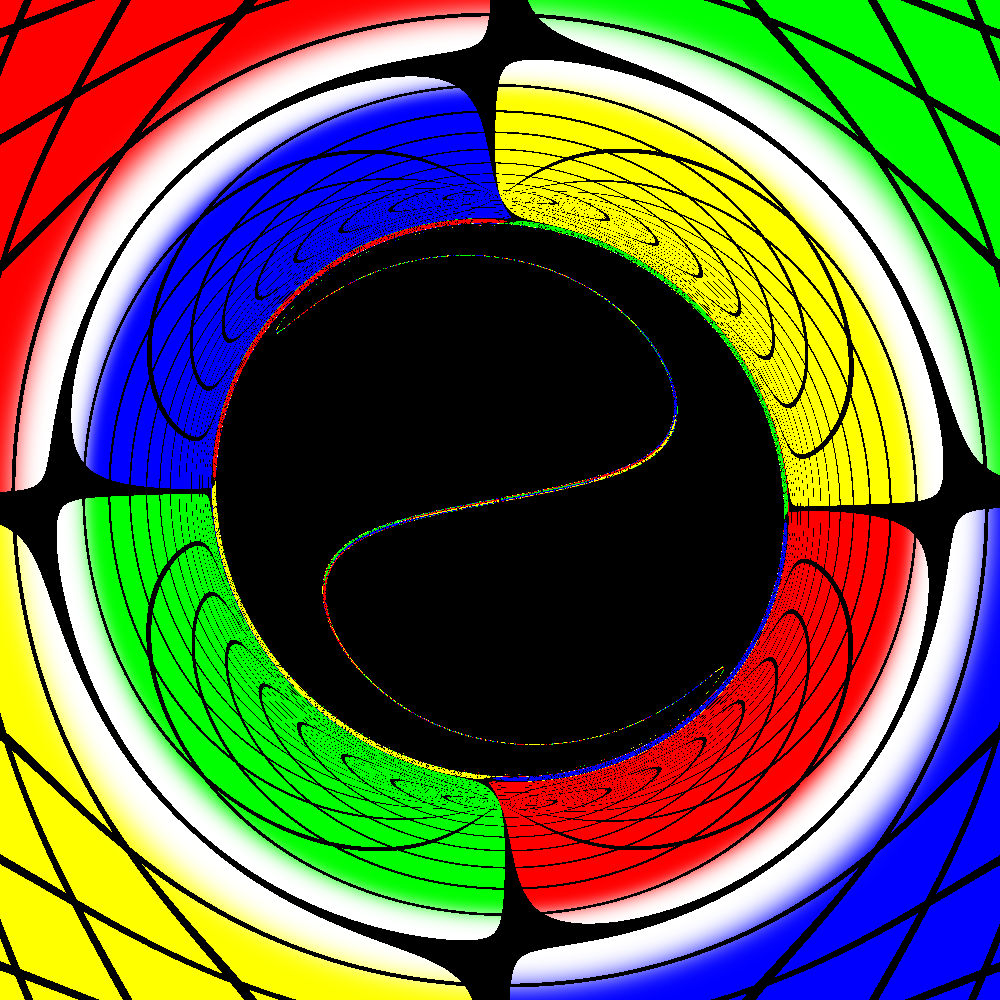}\includegraphics[width=0.25\textwidth]{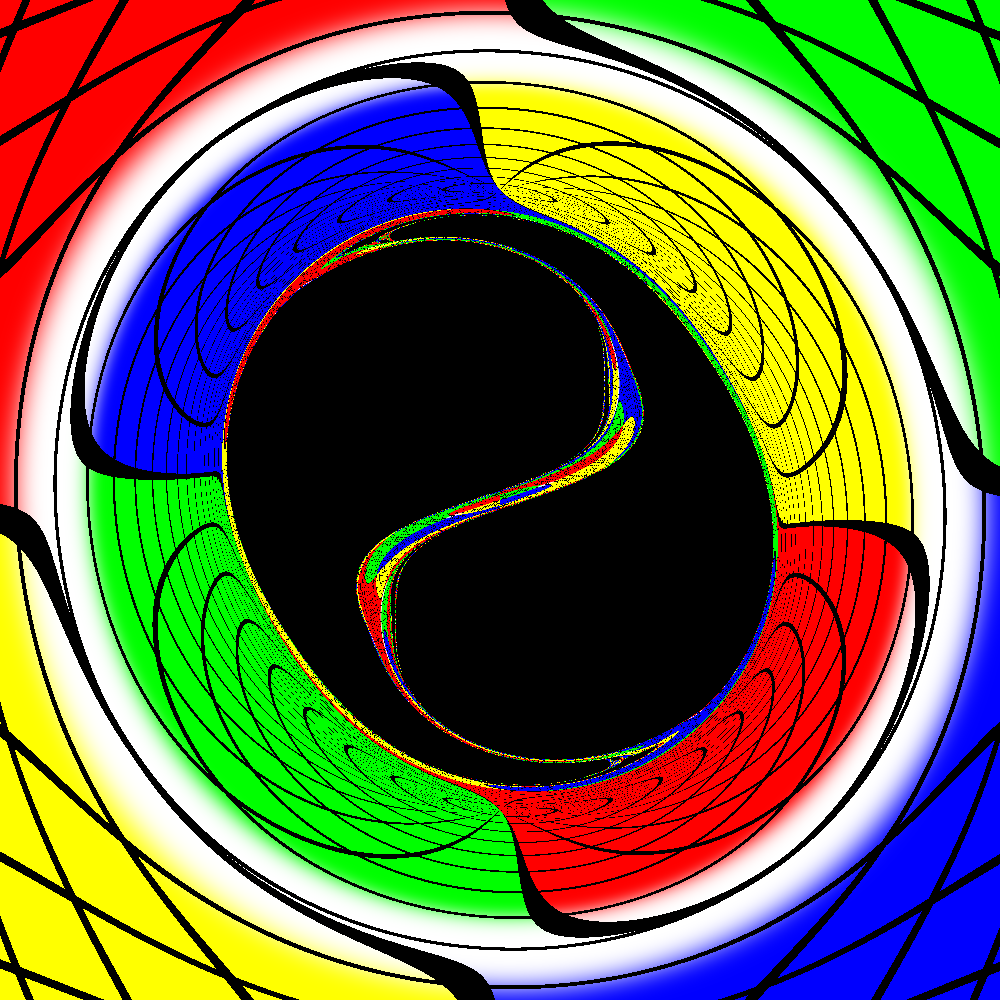}\includegraphics[width=0.25\textwidth]{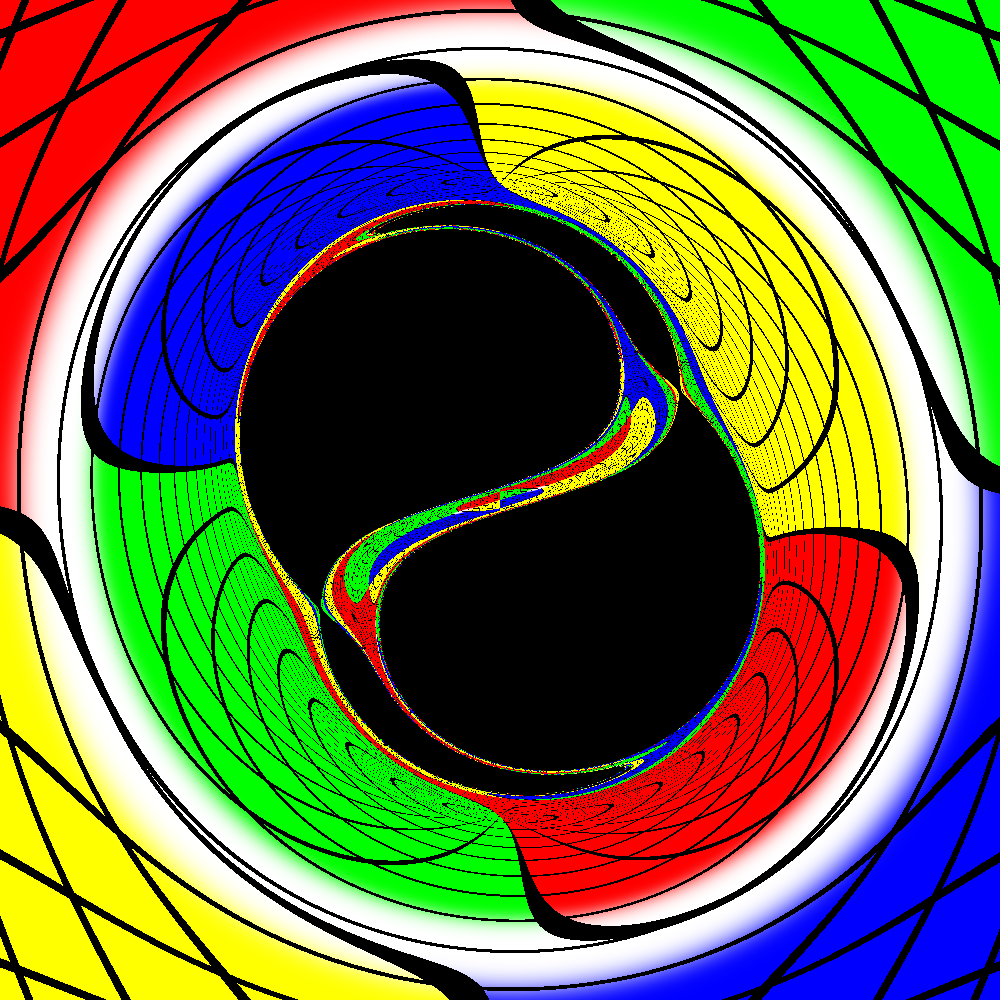}\\
\includegraphics[width=0.25\textwidth]{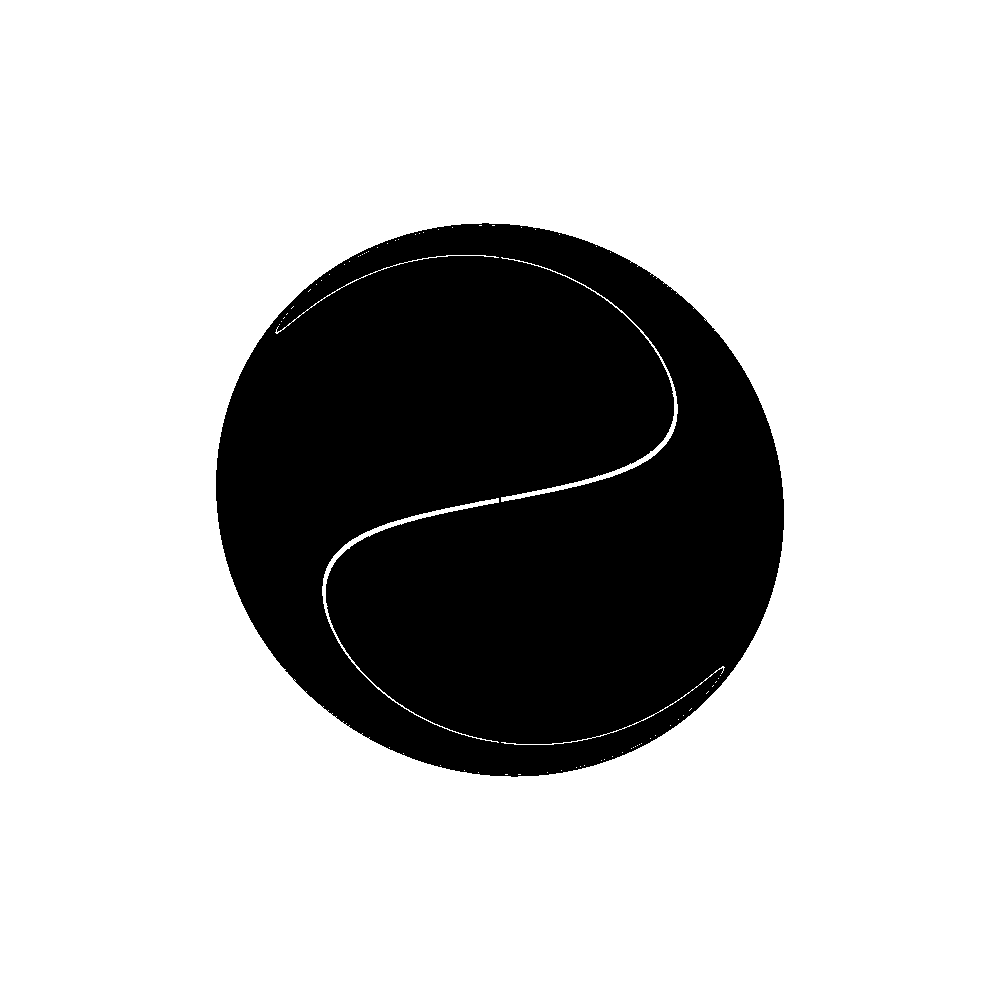}\includegraphics[width=0.25\textwidth]{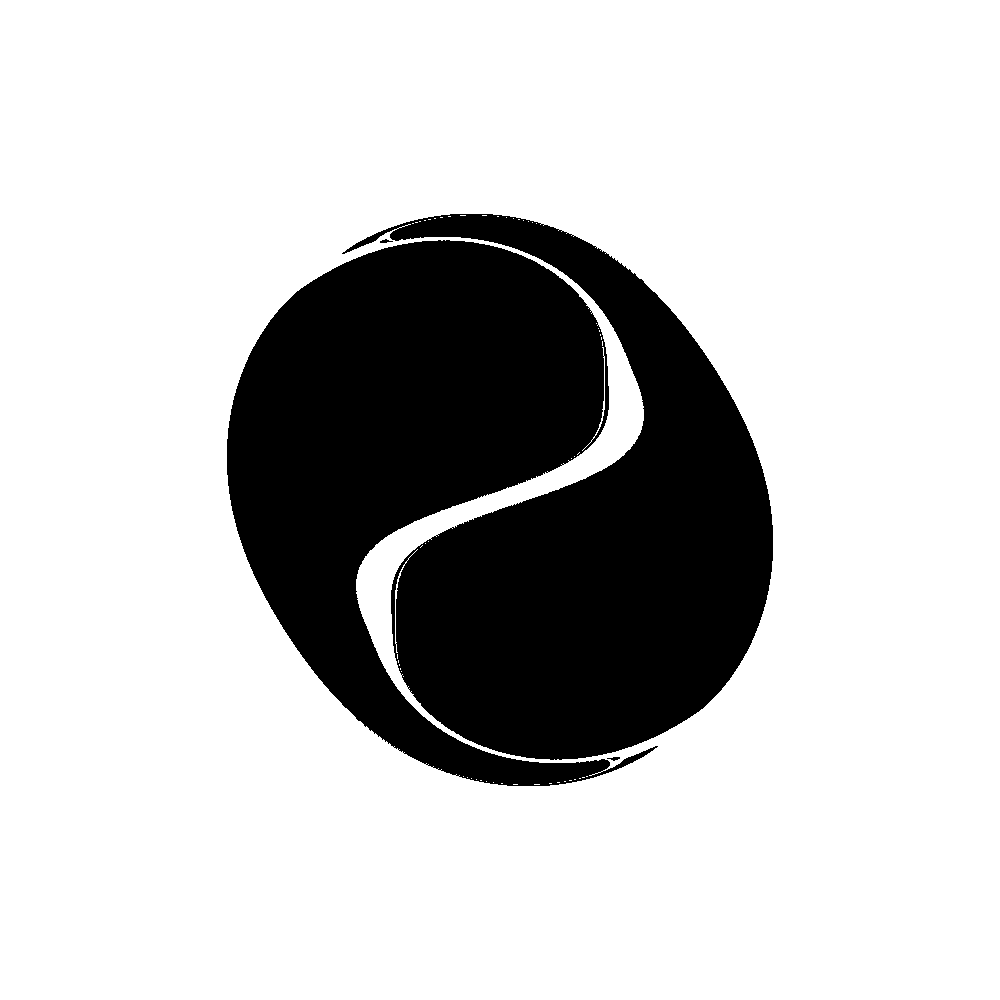}\includegraphics[width=0.25\textwidth]{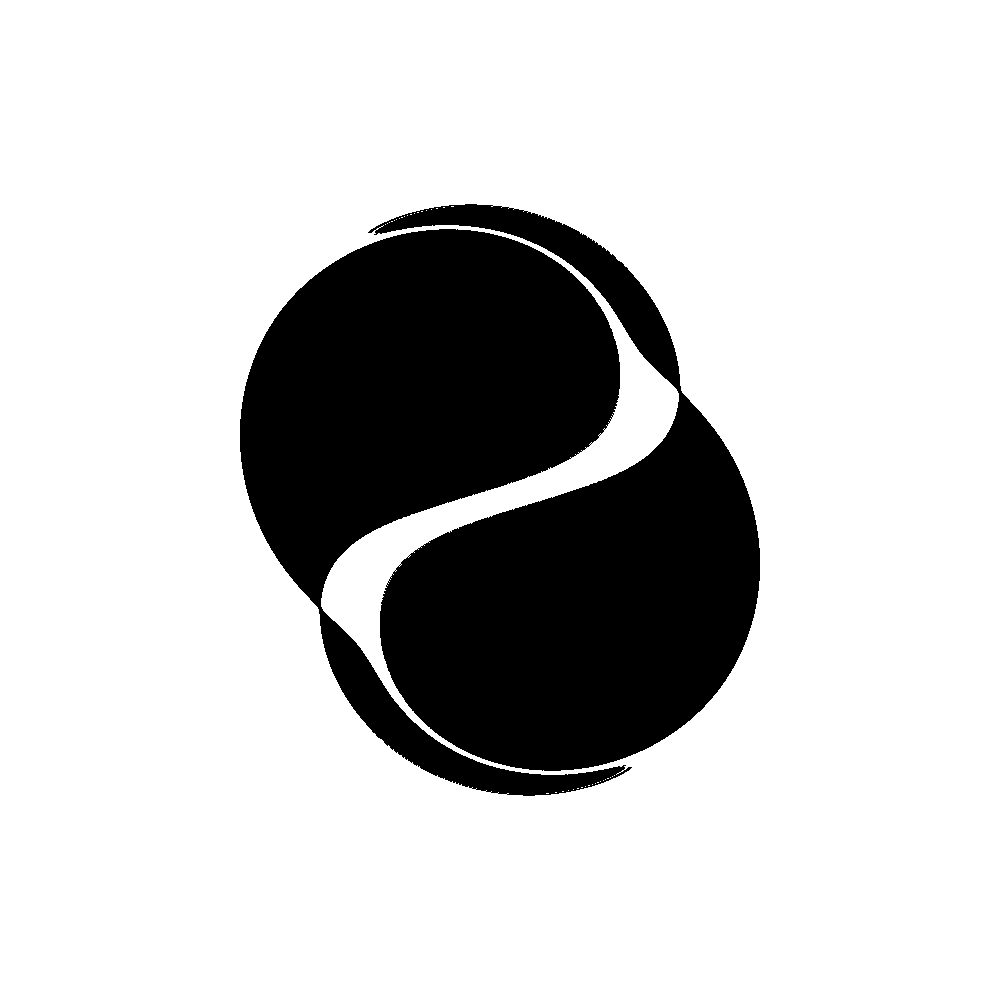}\\
\vspace{1.cm}
\includegraphics[width=0.25\textwidth]{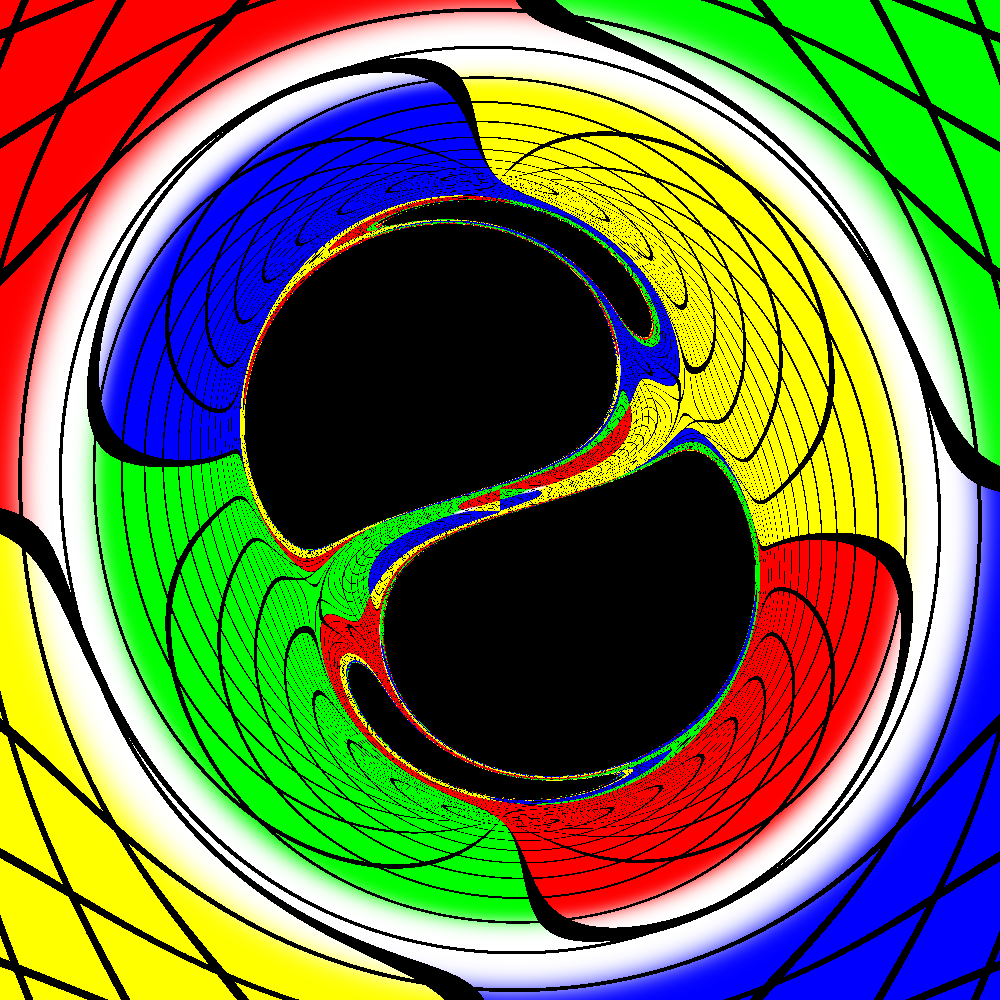}\includegraphics[width=0.25\textwidth]{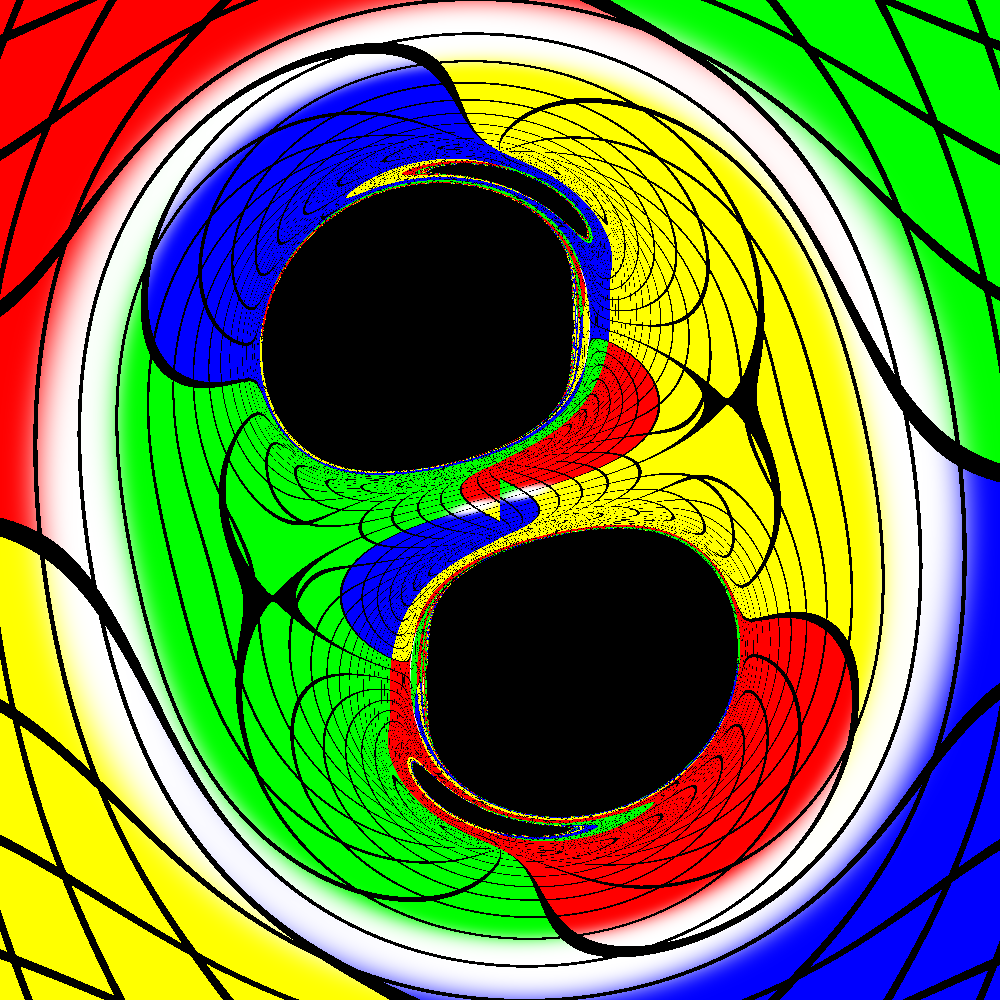}\includegraphics[width=0.25\textwidth]{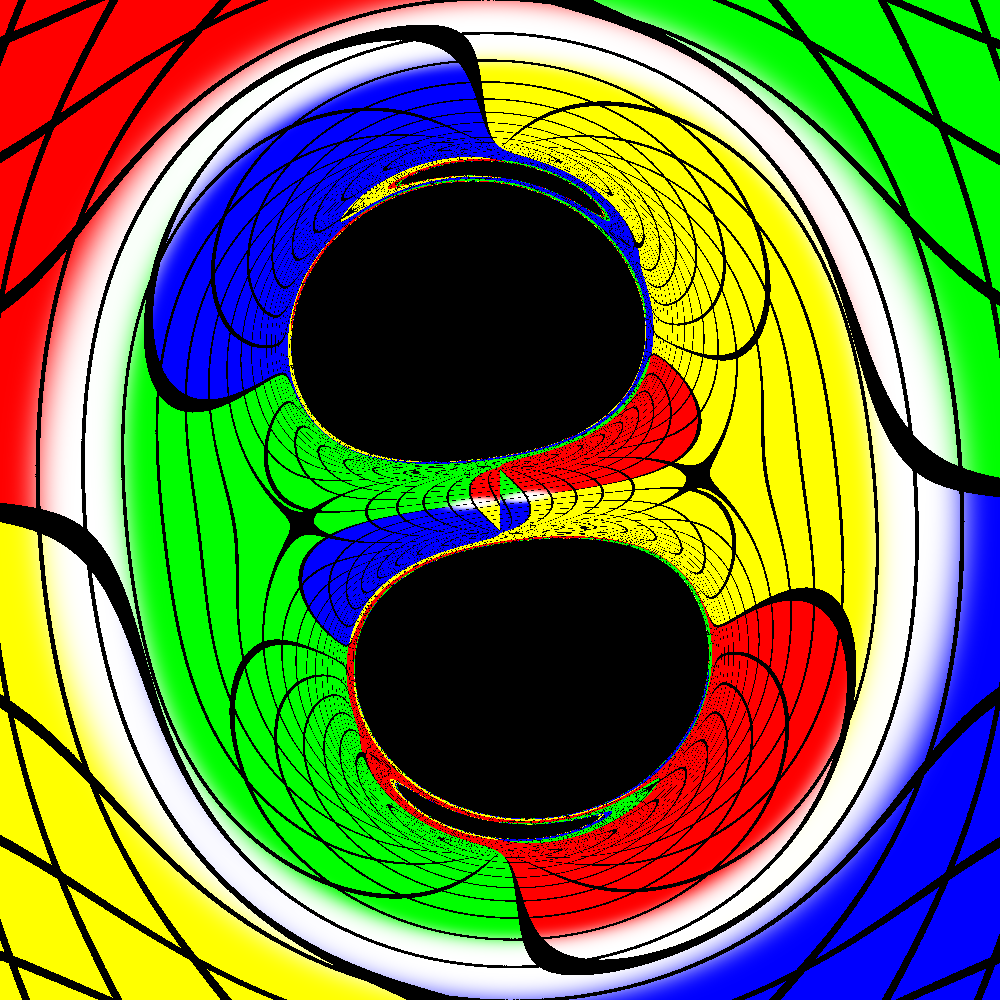}\\
\includegraphics[width=0.25\textwidth]{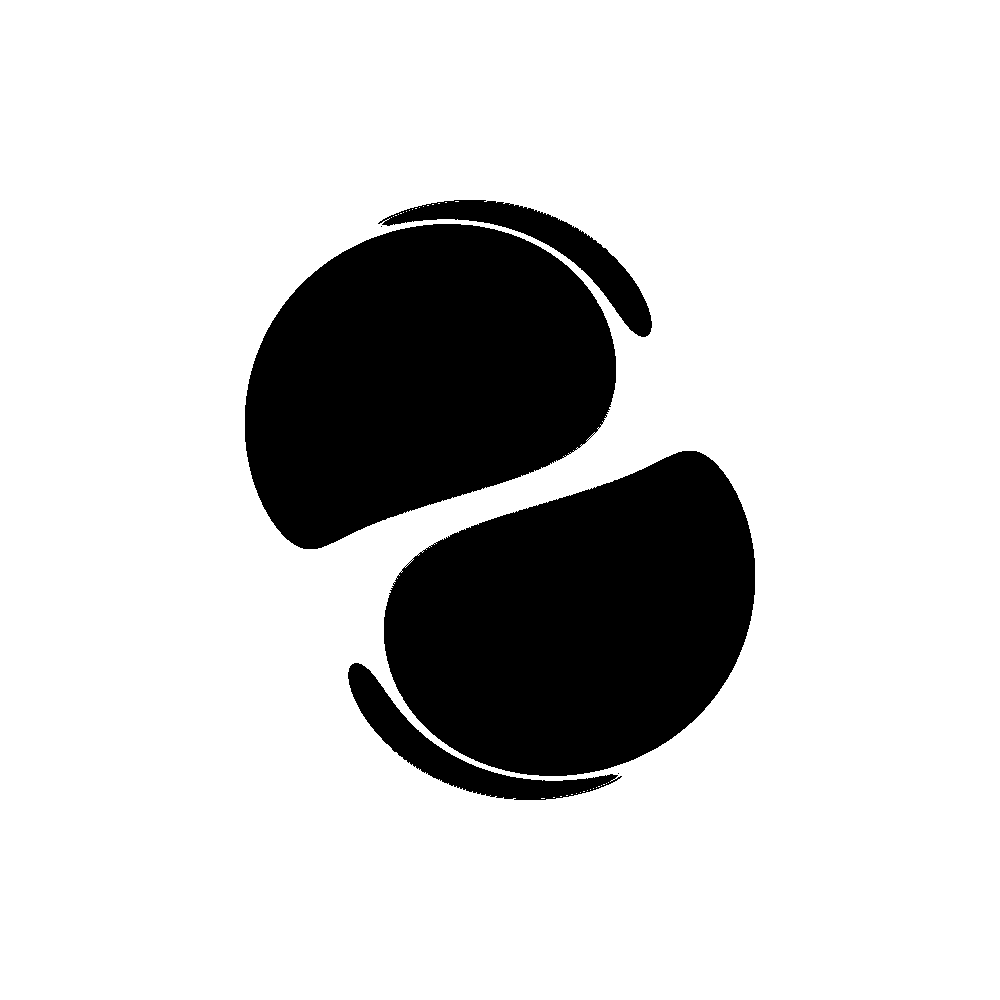}\includegraphics[width=0.25\textwidth]{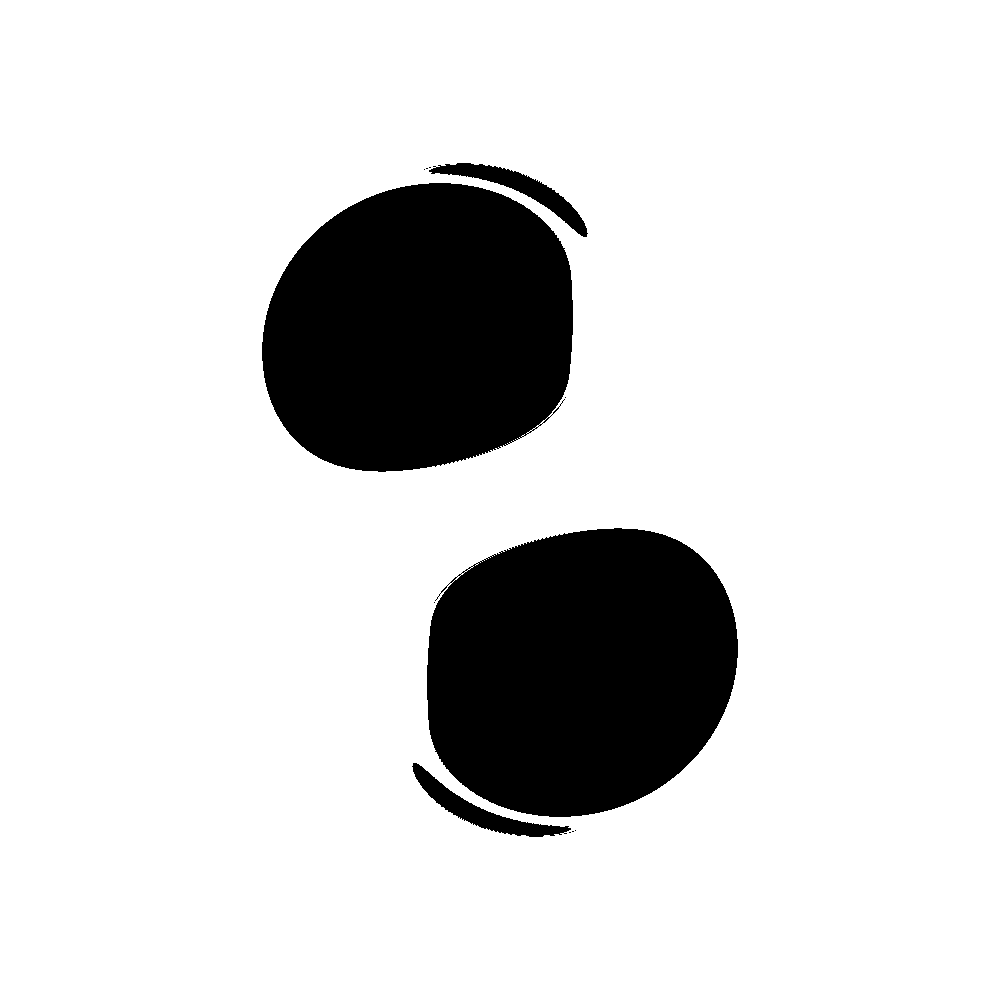}\includegraphics[width=0.25\textwidth]{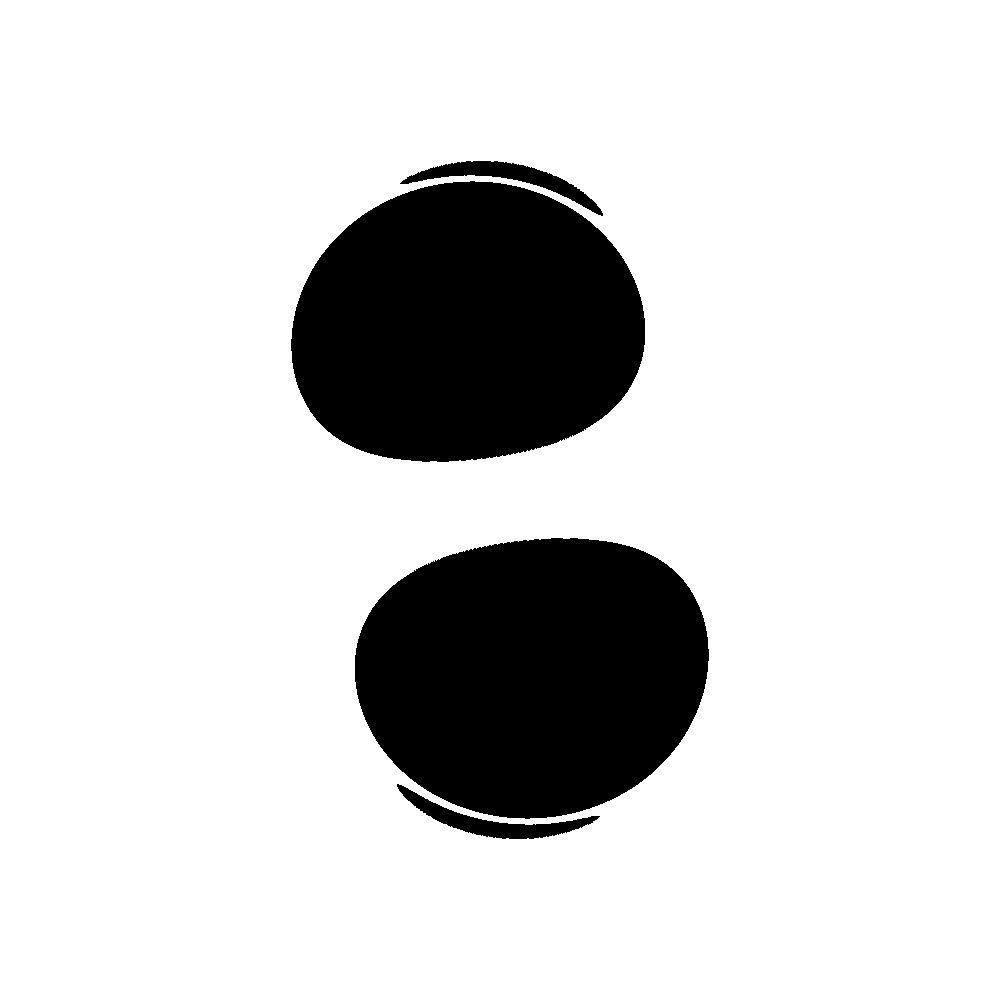}
\caption{\small Lensing of configurations 1 $\to$ 6 of Fig.~\ref{domain-2Kerr-odd} (from left to right and from top to bottom).}
\label{shadows-2Kerr-odd}
\end{center}
\end{figure}

The shadows and lensing of six solutions, marked in Fig.~\ref{domain-2Kerr-odd} with red dots, are displayed in Fig.~\ref{shadows-2Kerr-odd}.\footnote{Geodesics in the counter rotating Kerr-Newman solution were previously discussed in~\cite{Dubeibe:2016vhp}.}  The lensing and shadows appear to display a rotation effect, similar to that in Fig.~\ref{rotation}. However, despite the apparent similarities, both cases are quite different, with the anti-symmetry of the (odd) double-Kerr only giving the appearance of an image rotation. For instance, notice that the surface $z=0$ is not a totally geodesic sub-manifold, $i.e.$ a geodesic initially tangent to that plane can leave the latter, going up or down the plane depending on the sign of the geodesic angular momentum $L$. This effect together with anti-symmetric frame-dragging leaves the perception of a rotation at the level of the lensing. The image is stationary and not dynamical, in contrast to the quasi-static case in Fig.~\ref{rotation}. Another observation is that as $a\to \infty$ the shadows look increasingly Schwarzschild like, although there is still some shadow inner structure that quickly becomes imperceptible (see configuration 1 in Fig.~\ref{shadows-2Kerr-odd}). In addition, the shadow topology changes along the solutions, with configuration 3 displaying a shadow close to a topological transition.

\section{Discussion}

In this paper we have studied the effect of the orbital and intrinsic angular momentum in the lensing of light due to a BH binary, by using \textit{analytically} known solutions of General Relativity. In order to consider the effect of the orbital angular momentum, we have studied the double-Schwarzschild solution, which is static, under a quasi-static procedure that mimics an orbital rotation. The corresponding lensing is able to reproduce the main features of the shadows observed in dynamical binaries, obtained through a considerably more complex procedure which relies on producing fully non-linear numerical evolutions of BHs and performing ray tracing on top of these numerical evolutions. 

To observe the effect of the intrinsic spin of the BHs in the binary, we have considered the double-Kerr solution, which is stationary, for two particular cases: equal masses and equal or opposite spins. The lensing effects and shadow structure can be quite different in these two cases. In particular for the odd case,  an effect on the shadows similar, to some extent, to that of the orbital angular momentum can be observed, that can be traced back to the opposite dragging effects acting in the vicinity of the two BHs. 

One obvious further step would be to apply the quasi-static method of Section 2 to the double-Kerr stationary binaries of Section 3. Whereas the procedure should be straightforward, the involved nature of the double-Kerr metric makes it cumbersome. We expect the end result for the shadows to be a superposition of the orbital effect seen in Section 2 with the corresponding intrinsic spin effect seen in Section 3.

\section*{Acknowledgements}

We would like to thank the authors of Ref.~\cite{2015CQGra..32f5002B} for their permission to use the left panel of Fig.~\ref{bohn}. We would also like to thank E. Radu for discussions. P.C. is supported by Grant No. PD/BD/114071/2015 under the FCT-IDPASC Portugal Ph.D. program. 
C.H. acknowledges funding from the FCT-IF programme. This work was partially supported by the H2020-MSCA-RISE-2015 Grant No. StronGrHEP-690904,
the H2020-MSCA-RISE-2017 Grant No. FunFiCO-777740 and by the CIDMA project UID/MAT/04106/2013. The authors would like to acknowledge networking support by the
COST Action CA16104. The work of MJR was supported by the Max Planck Gesellschaft through the Gravitation and Black Hole Theory Independent Research Group and by NSF grant PHY-1707571 at Utah State University.

\bibliography{Ref}{}  
\bibliographystyle{ieeetr}

\end{document}